{}%disable hyper at arxiv
{}%use pdflatex at arxiv

\documentclass[12pt,a4paper]{article}

\newif\ifpublic\publictrue
\newif\ifniklas\niklastrue

\usepackage{amsmath,amssymb}
\ifpublic\else\usepackage{showkeys}\fi
\usepackage[bookmarks=true,hyperfigures=true]{hyperref}
\usepackage{graphicx}
\usepackage[nosort]{cite}
\usepackage[bulletsep]{collref}
\ifniklas

\def\showkeysrefformat#1{{\normalfont\tiny\ttfamily#1}}
\makeatletter
\def\SK@@ref#1>#2\SK@{%
 {\@inlabelfalse\leavevmode\vbox to\z@{%
 \vss\SK@refcolor\rlap{\vrule\raise .75em%
  \hbox{\showkeysrefformat{#2}}}}}}
\makeatother
\fi

\usepackage[a4paper,text={160mm,247mm},centering]{geometry}
\allowdisplaybreaks[3]

\numberwithin{equation}{section}

\usepackage[font=small,labelfont=bf,width=0.85\textwidth]{caption}

\expandafter\def\expandafter\bfseries\expandafter{\bfseries\ifmmode\else\boldmath\fi}
\expandafter\def\expandafter\mdseries\expandafter{\mdseries\ifmmode\else\unboldmath\fi}
\expandafter\def\expandafter\normalfont\expandafter{\normalfont\ifmmode\else\unboldmath\fi}

\RequirePackage{verbatim}

\makeatletter
\newwrite\bibinl@out
\newenvironment{bibtex}[1][\jobname]{%
  \immediate\openout\bibinl@out #1.bib
  \immediate\write\bibinl@out{\@percentchar generated from `\jobname' starting line \the\inputlineno^^J}%
  \def\verbatim@processline{\immediate\write\bibinl@out{\the\verbatim@line}}%
  \@bsphack\let\do\@makeother\dospecials\catcode`\^^M\active\verbatim@start
}%
{\immediate\closeout\bibinl@out\@esphack}
\makeatother

\RequirePackage{verbatim}

\makeatletter
\newwrite\mpi@out
\def\mpi@write#1{\immediate\write\mpi@out{#1}}
\immediate\openout\mpi@out\jobname.mp
\mpi@write{\@percentchar generated from `\jobname'}%
\AtEndDocument{\immediate\closeout\mpi@out}

\newcommand{\mpi@putlineno}{%
  \mpi@write{\@percentchar---------------------------------------}%
  \mpi@write{\@percentchar l.\the\inputlineno}%
}

\newcommand{\mpi@verbatim}{
  \@bsphack
  \let\do\@makeother\dospecials
  \catcode`\^^M\active
  \def\verbatim@processline{\mpi@write{\the\verbatim@line}}%
  \verbatim@start
}

\newenvironment{mpostcmd}{%
  \mpi@putlineno%
  \mpi@verbatim%
}%
{\mpi@write{}\@esphack}

{\mpi@write{endfig;}\@esphack}

\newcommand{\includegraphicsex}[2][]{%
  \xdef\mpi@tmp{#2}%
  \IfFileExists{\mpi@tmp}%
    {\includegraphics[#1]{\mpi@tmp}}%
    {\textbf{??}\typeout{file \mpi@tmp{} missing}}%
}

\makeatother

\usepackage{fixmath}

\ifx\genfrac\sdflkaj\else\fi
\newcommand{\sfrac}[2]{{\textstyle\frac{#1}{#2}}}
\newcommand{\half}{\sfrac{1}{2}}

\newcommand{\quarter}{\sfrac{1}{4}}
\newcommand{\gammafn}{\mathrm{\Gamma}}

\newcommand{\hopf}[1]{\mathrm{#1}}
\newcommand{\yang}{\hopf{Y}}

\newcommand{\alg}[1]{\mathfrak{#1}}
\newcommand{\grp}[1]{\mathrm{#1}}

\newcommand{\copro}{\mathrm{\Delta}}

\newcommand{\bigbrk}[1]{\bigl(#1\bigr)}

\newcommand{\sprods}[2]{\langle#1#2\rangle}

\newcommand{\cprods}[2]{[#1#2]}

\newcommand{\superN}{\mathcal{N}}

\newcommand{\nn}{\nonumber}
\newcommand{\nln}{\nonumber\\}

\def\[{\begin{equation}}
\def\]{\end{equation}}

\providecommand{\href}[2]{#2}
\newcommand{\arxivlink}[1]{\href{http://arxiv.org/abs/#1}{arxiv:#1}}

\makeatletter
\def\mr@ignsp#1 {\ifx\:#1\@empty\else #1\expandafter\mr@ignsp\fi}%
\newcommand{\multiref}[1]{\begingroup%\let\protect\string%
\xdef\mr@no@sparg{\expandafter\mr@ignsp#1 \: }%
\def\mr@comma{}%
\@for\mr@refs:=\mr@no@sparg\do{\mr@comma\def\mr@comma{,}\ref{\mr@refs}}%
\endgroup}
\renewcommand{\eqref}[1]{(\multiref{#1})}
\makeatother

\makeatletter
\newcommand{\namedref}[2]{\hyperref[#2]{#1~\ref*{#2}}}
\newcommand{\secref}{\@ifstar{\namedref{Section}}{\namedref{sec.}}}
\newcommand{\subsecref}{\@ifstar{\namedref{Subsection}}{\namedref{subsec.}}}
\newcommand{\appref}{\@ifstar{\namedref{Appendix}}{\namedref{app.}}}
\newcommand{\tabref}{\@ifstar{\namedref{Table}}{\namedref{tab.}}}
\newcommand{\figref}{\@ifstar{\namedref{Figure}}{\namedref{fig.}}}
\makeatother

\providecommand{\hypersetup}[1]{}
\providecommand{\texorpdfstring}[2]{#1}

\hypersetup{plainpages=false}
\hypersetup{pdfpagemode=UseNone}
\hypersetup{bookmarksnumbered=true}
\hypersetup{pdfstartview=FitH}
\hypersetup{colorlinks=false}
\hypersetup{citebordercolor={.5 1 .5}}
\hypersetup{urlbordercolor={.5 1 1}}
\hypersetup{linkbordercolor={1 .7 .7}}
\makeatletter
\let\@keywords\@empty
\let\@subject\@empty
\providecommand{\keywords}[1]{\gdef\@keywords{#1}}
\providecommand{\subject}[1]{\gdef\@subject{#1}}
\def\thetitle{\@title}
\def\theauthor{\@author}
\def\thesubject{\@subject}
\def\thedate{\@date}
\def\thekeywords{\@keywords}
\makeatother
\AtBeginDocument{
\hypersetup{pdftitle={\thetitle}}%
\hypersetup{pdfauthor={\theauthor}}%
\hypersetup{pdfsubject={\thesubject}}%
\hypersetup{pdfkeywords={\thekeywords}}%
}

\makeatletter
\newsavebox{\apb@box}\newlength{\apb@width}
\newcommand{\autoparbox}[2][c]{\sbox{\apb@box}{#2}%
 \settowidth{\apb@width}{\usebox{\apb@box}}%
 \parbox[#1]{\apb@width}{\usebox{\apb@box}}}

\newcommand{\includegraphicsboxex}[2][]{\autoparbox{\includegraphicsex[#1]{#2}}}
\makeatother

\newif\ifmrnote 
\mrnotetrue
 
\newcommand{\zvar}{c}
\usepackage[active]{srcltx}
\newcommand{\why}[1]{\mathfrak{#1}}

\newif\ifjbnote 
\jbnotetrue
 
\newcommand{\eqnref}[1]{eq.~(\ref{#1})}

\begin{mpostcmd}
prologues:=3;

verbatimtex 
\documentclass{article}
\begin{document}
etex

\end{mpostcmd}

\begin{mpostcmd}
verbatimtex 
\end{document}
etex

end;
\end{mpostcmd}

%%%%%%%%%%%%%%%%%%%%%%%%%%%%%%%%%%%%%%%%%%%%%%%%%%%%%%%%%%%%%%%%%%%%%%%%%%%%%%%%
%%%%%%%%%%%%%%%%%%%%%%%%%%%%%%%%%%%%%%%%%%%%%%%%%%%%%%%%%%%%%%%%%%%%%%%%%%%%%%%%
%\title{\texorpdfstring{$\xi^2$}{\xi^2}}
\title{On Yangian-invariant regularisation of deformed on-shell diagrams in
  \texorpdfstring{$\mathcal{N}=4$}{N=4} super-Yang--Mills theory}%
\author{Niklas Beisert, Johannes Broedel and Matteo Rosso}

%%%%%%%%%%%%%%%%%%%%%%%%%%%%%%%%%%%%%%%%%%%%%%%%%%%%%%%%%%%%%%%%%%%%%%%%%%%%%%%%
\begin{document}

\iftrue

\pdfbookmark[1]{Title Page}{title} \thispagestyle{empty}

\begingroup\raggedleft\footnotesize\ttfamily
\arxivlink{1401.7274}
\par\endgroup

\vspace*{2cm}
\begin{center}%
  \begingroup\Large\bfseries\thetitle\par\endgroup
  \vspace{1cm}

\begingroup\scshape\theauthor\par\endgroup
\vspace{5mm}%

\begingroup\itshape
Institut f\"ur Theoretische Physik,\\
Eidgen\"ossische Technische Hochschule Z\"urich\\
Wolfgang-Pauli-Strasse 27, 8093 Z\"urich, Switzerland
\par\endgroup
\vspace{5mm}

\begingroup\ttfamily
\verb+{+nbeisert,jbroedel,mrosso\verb+}+@itp.phys.ethz.ch
\par\endgroup

\vfill

\textbf{Abstract}\vspace{5mm}

\begin{minipage}{12.7cm}
  We investigate Yangian invariance of deformed on-shell diagrams with $D=4$,
  $\superN=4$ superconformal symmetry.  We find that invariance implies a direct
  relationship between the deformation parameters and the permutation associated
  to the on-shell graph.  We analyse the connection with deformations of
  scattering amplitudes in $\superN=4$ super-Yang--Mills theory and the
  possibility of using the deformation parameters as a regulator preserving
  Yangian invariance. A study of higher-point tree and loop graphs suggests that
  manifest Yangian invariance of the amplitude requires trivial deformation
  parameters.
\end{minipage}

\vspace*{4cm}

\end{center}

\newpage

\fi

\tableofcontents
\vspace{0.8cm}
\hrule height 0.75pt
\vspace{1cm}

%%%%%%%%%%%%%%%%%%%%%%%%%%%%%%%%%%%%%%%%%%%%%%%%%%%%%%%%%%%%%%%%%%%%%%%%%%%%%%%%
%%%%%%%%%%%%%%%%%%%%%%%%%%%%%%%%%%%%%%%%%%%%%%%%%%%%%%%%%%%%%%%%%%%%%%%%%%%%%%%%

\section{Introduction}

One of the most striking features of the $\superN=4$ super-Yang--Mills (sYM)
theory is that its tree-level S-matrix is invariant under the
infinite-dimensional Yangian algebra $\yang[\alg{psu}(2,2|4)]$
\cite{Drummond:2009fd}, see also
\cite{Roiban:2011zz,Beisert:2010jq}. This symmetry arises as
the closure of the ordinary superconformal symmetry together with the hidden
dual superconformal symmetry the theory possesses~\cite{Drummond:2008vq}. The
existence of this infinite-dimensional symmetry algebra is another important
evidence in support of the conjectured integrability of the planar sector of
the theory, see \cite{Beisert:2010jr}.  It is however well known that IR
divergences break the (dual) superconformal symmetry of scattering amplitudes
at loop level to some extent \cite{Drummond:2008vq}.

In ref.~\cite{ArkaniHamed:2012nw}, the authors proposed an interesting
construction relating Yangian invariants in $\superN=4$ sYM to so-called
on-shell graphs (or on-shell diagram). These graphs are associated with
integrals over suitably defined subspaces of Grassmannian manifolds and provide
a direct link to the Grassmannian formulation of scattering amplitudes,
introduced and studied in 
refs.~\cite{ArkaniHamed:2009dn,Drummond:2010qh,Drummond:2010uq,ArkaniHamed:2010kv,ArkaniHamed:2010gh}.

In refs.\cite{Ferro:2012xw,Ferro:2013dga}, the authors introduced
\emph{deformed} on-shell graphs, where the deformation consists of the shift of
the helicities of the external legs by a complex value. In this article we
focus on the conditions for manifest Yangian invariance for such deformed
graphs.

A topic which is worth investigating in this framework is the relation between
deformed on-shell diagrams and scattering amplitudes, a relation which is
currently unclear. We will study the possibility of constructing such a
correspondence in the context of the tree-level Britto--Cachazo--Feng--Witten
(BCFW) recursion relations and their supersymmetric
extension~\cite{Britto:2004ap,Britto:2005fq,ArkaniHamed:2008gz,Brandhuber:2008pf}.
Our starting point will be the formulation of the BCFW recursion relations in
terms of undeformed on-shell graphs. We will subsequently focus on one simple
example, the six-point NMHV amplitude, and study the possibility of
constructing a manifestly Yangian-invariant deformed amplitude as a sum of
deformed Yangian invariants.

In refs.\ \cite{Ferro:2012xw,Ferro:2013dga} it was also proposed that the
deformation parameters could be used as regulators for loop amplitudes.  As an
example, the four-point one-loop amplitude in $\superN=4$ sYM was explicitly
computed for a particular choice of deformation which happens to break 
Yangian invariance. 
We will address the question of the compatibility of this regulating method
with Yangian invariance by computing the four-point one-loop deformed
on-shell graph with a Yangian-preserving set of deformation parameters.

The paper is organised as follows: In \secref{sec:yanginv} we review the
concept of Yangian invariance, we define undeformed and deformed on-shell
graphs and discuss the conditions these deformed on-shell graphs have to
satisfy in order to be Yangian invariant. \secref*{sec:amplitudes} deals with
the translation from Yangian invariants represented by deformed on-shell graphs
to scattering amplitudes. We will focus in particular on tree-level
$\mathrm{NMHV}$ amplitudes as well as the four-point one-loop amplitude in
$\superN=4$ sYM and discuss their compatibility with the deformation of
external helicities. \secref*{sec:conclusions} contains a summary of the
results.

%%%%%%%%%%%%%%%%%%%%%%%%%%%%%%%%%%%%%%%%%%%%%%%%%%%%%%%%%%%%%%%%%%%%%%%%%%%%%%%%
%%%%%%%%%%%%%%%%%%%%%%%%%%%%%%%%%%%%%%%%%%%%%%%%%%%%%%%%%%%%%%%%%%%%%%%%%%%%%%%%

\section{Yangian invariance}
\label{sec:yanginv}

A mechanism to build Yangian invariants starting from three-point vertices in
$\superN=4 $ sYM and producing so-called on-shell diagrams was developed in
ref.~\cite{ArkaniHamed:2012nw}.  Yangian invariance, however, allows for more
general building blocks: in this chapter we will use \emph{deformed}
three-point vertices (developed in ref.~\cite{Ferro:2012xw,Ferro:2013dga}) and
investigate the properties of the resulting deformed on-shell diagrams. The
relation to scattering amplitudes in $\superN=4$ sYM theory will be discussed
in~\secref{sec:amplitudes} below.

%%%%%%%%%%%%%%%%%%%%%%%%%%%%%%%%%%%%%%%%%%%%%%%%%%%%%%%%%%%%%%%%%%%%%%%%%%%%%%%%

\subsection{Yangian symmetry and representation on spinor variables}

\label{YangianIntroduction}

The Yangian algebra $\yang:=\yang{[\alg{g}]}$ is a quantum algebra based on (half of)
the affine extension of the Lie algebra%
\footnote{In what follows, we will not
  distinguish between Lie algebras and Lie superalgebras. When considering Lie
  superalgebras, all expressions should be understood with the appropriate
  graded symmetrisation and factors of $(-1)^{\mathrm{deg}(\dots)}$.} $\alg{g}$.
  In addition to the generators $\alg{J}^a$ of the Lie algebra $\alg{g}$ obeying
  the usual Lie-algebra relations
\begin{align}
  [\alg{J}^a,\alg{J}^b]=f_c^{ab}\,\alg{J}^c,
\end{align}
there are level-one generators $\widehat{\alg{J}}^a$ satisfying
\begin{align}
  [\alg{J}^a,\widehat{\alg{J}}^b]=f_c^{a b}\,\widehat{\alg{J}}^c\,,
\end{align}
and the Serre relations
\begin{align}
  \label{eq:Serre}
  \Bigl[\big[\alg{J}^a,\widehat{\alg{J}}^b],\widehat{\alg{J}}^c\Bigr]+
  \Bigl[\big[\alg{J}^b,\widehat{\alg{J}}^c],\widehat{\alg{J}}^a\Bigr]+
  \Bigl[\big[\alg{J}^c,\widehat{\alg{J}}^a],\widehat{\alg{J}}^b\Bigr]=
  f_d^{a g}f_e^{b h}f_f^{c i}f_{g h i}\,\alg{J}^{\lbrace
    d}\alg{J}^e\alg{J}^{f \rbrace}\,,
\end{align}
where $\lbrace\ldots\rbrace$ means graded symmetrisation of indices.
The complete Yangian algebra $\yang[\alg{g}]$ is obtained by successively
commuting level-zero and level-one generators.  The coproduct $\copro \,:\,
\yang\to \yang\otimes \yang$ of the Yangian algebra is defined on the level-zero
and level-one generators as
\[
\label{eq:coproduct_def}
\copro(\alg{J}^a):= \alg{J}^a\otimes 1 + 1 \otimes \alg{J}^a\quad\text{and}\quad
\copro(\widehat{\alg{J}}^a):=\widehat{\alg{J}}^a\otimes 1 + 1 \otimes
\widehat{\alg{J}}^a + f^{a}_{\,b c} \,\alg{J}^b\otimes \alg{J}^c,
\]
respectively. 

In this article, we will consider $\yang{[\alg{psu}(2,2|4)]}$, the Yangian of the
maximal superconformal algebra in four dimensions. A convenient representation
of $\alg{psu}(2,2|4)$ acts on functions defined on the on-shell superspace with
spinor-helicity variables $\lambda^\alpha,\tilde{\lambda}^{\dot{\alpha}}$ and
Grassmann variables $\tilde{\eta}^A$. Here, dotted and undotted Greek spinor
indices take the values $1$ and $2$, and capital Latin indices correspond to a
four-dimensional fundamental representation of $\grp{SU}(4)$.

The spinors $\lambda$ and $\tilde{\lambda}$ are related to the massless momentum
vector via $p_\mu=\sigma_{\mu\,\alpha\dot{\alpha}}\lambda^\alpha
\tilde{\lambda}^{\dot{\alpha}}$, where $\sigma_{\mu\alpha\dot{\alpha}}$ are
generalised Pauli matrices which have to be chosen according to the spacetime
signature.  Here we work with complexified momenta, which amounts to treating
$\lambda$ and $\tilde{\lambda}$ as independent complex
variables~\cite{Mason:1991rf}. 
For a fixed external momentum $p_\mu$ the spinors $\lambda$ and
$\tilde{\lambda}$ are defined up to a rescaling
\[
  \lambda\to \alpha \lambda,\qquad\tilde{\lambda}\to \alpha^{-1} \tilde{\lambda}\,,
\label{eq:alphascaling}
\]
where $\alpha$ is an arbitrary nonzero complex number. In the supersymmetric
part of the on-shell space, the Grassmann variables have to scale as 
\[
  \tilde{\eta}\to \alpha^{-1} \tilde{\eta}
\label{eq:etascaling}
\]
for consistency. Imposing reality of momenta in signature $(1,3)$ amounts to equating
$\tilde{\lambda}=\pm \bar{\lambda}$, where the sign depends on the sign of the
energy. This implies that the freedom of rescaling $\lambda,\tilde{\lambda}$
for real momenta is restricted to multiplication by a phase.

In this representation, the generators of $\alg{psu}(2,2|4)$ read (see
ref.~\cite{Witten:2003nn})
\[
\label{eq:superconformal_onshell_superspace}
\begin{aligned}
  \why{L}^\alpha_{\,\beta} & = \lambda^\alpha\frac{\partial}{\partial
    \lambda^\beta}-\half \delta^\alpha_{\,\beta}
  \lambda^\gamma\frac{\partial}{\partial \lambda^\gamma},&
  \bar{\why{L}}^{\dot{\alpha}}_{\,\dot{\beta}} & =
  \lambda^{\dot{\alpha}}\frac{\partial}{\partial \lambda^{\dot{\beta}}}-\half
  \delta^{\dot{\alpha}}_{\,\dot{\beta}}
  \lambda^\gamma\frac{\partial}{\partial \lambda^\gamma},  \\\
  \why{Q}^{A\alpha} & = \lambda^\alpha \tilde{\eta}^A,&
  \bar{\why{Q}}_{A}^{\dot{\alpha}} & = \tilde{\lambda}^{\dot{\alpha}}
  \frac{\partial }{\partial \tilde{\eta}^A}, \\
  \why{S}_{A\alpha} & = \frac{\partial^2}{\partial\lambda^\alpha \partial
    \tilde{\eta}^A},&
  \bar{\why{S}}^A_{\dot{\alpha}} & =
  \tilde{\eta}^A\frac{\partial}{\partial \tilde{\lambda}^{\dot{\alpha}}}, \\
  \why{R}^A_{\,B} & = \tilde{\eta}^A\frac{\partial}{\partial\tilde{\eta}^B} -
  \quarter \delta^{A}_{\,B}\tilde{\eta}^C\frac{\partial}{\partial \tilde{\eta}^C},&
  \why{D} & = \half \lambda^\gamma\frac{\partial}{\partial \lambda^\gamma} +
  \half \tilde{\lambda}^{\dot{\gamma}} \frac{\partial}{\partial \tilde{\lambda}^{\dot{\gamma}}}+1,\\
  \why{P}^{\alpha\dot{\alpha}} & =\lambda^\alpha \tilde{\lambda}^{\dot{\alpha}},
  & \why{K}_{\alpha\dot{\alpha}} & =\frac{\partial^2}{\partial
    \lambda^\alpha \partial\tilde{\lambda}^{\dot{\alpha}}}.
\end{aligned}
\]
The above representation of the Lie superalgebra $\alg{psu}(2,2|4)$ can be lifted
to an evaluation representation of the Yangian $\yang{[\alg{psu}(2,2|4)]}$, for
which the level-one generators act as $\widehat{\alg{J}}\simeq u \alg{J}$. Here
$u$ is called the evaluation parameter of the representation. In the evaluation
representation, the left-hand-side of the Serre relation \eqnref{eq:Serre}, reduces
to the usual Jacobi identity. Accordingly, the right hand side can be shown to vanish. 

With the evaluation representation on disposal, one can construct the action of
the Yangian algebra on functions depending on several variables in the on-shell
superspace. This action can be obtained from the maximally iterated coproduct  
\[
\label{eq:nth_tensorproduct}
\copro^{n-1}(\widehat{\alg{J}}^a) = \sum_{k=1}^n \widehat{\alg{J}}^a_{\,k} +
f^a_{\,b c}\sum_{1\leq i < j \leq n}\,\alg{J}^b_i\,\alg{J}^c_j = \sum_{k=1}^n
u_k\alg{J}^a_{\,k} + f^a_{\,b c}\sum_{1\leq i < j \leq
  n}\,\alg{J}^b_i\,\alg{J}^c_j,
\]
where the first equality holds due to the definition of the coproduct
\eqnref{eq:coproduct_def}, whereas the second equality applies to evaluation
representations.  The evaluation parameters $u_k$ can (and will) be different
for each element of the tensor product.

In addition to the generators defined in
\eqnref{eq:superconformal_onshell_superspace} there is another important quantity
for the discussion to follow: the central charge operator $\alg{C}$ that appears
in the anticommutator of $\why{Q},\why{S}$
\[
\bigl\{\why{Q}^{A\alpha},\why{S}_{B\beta}\bigr\} = \delta^A_B \why{L}^\alpha_{\,\beta}
- \delta^{\alpha}_\beta \why{R}^A_{\,B}+\tfrac{1}{2} \delta^A_B
\delta^\alpha_\beta \bigl[\why{D} + \tfrac{1}{2}\why{C}\bigr].
\]
It is represented by
\[
\label{eq:central_charge_op}
\alg{C} := \lambda^\alpha \frac{\partial}{\partial \lambda^\alpha} -
\tilde{\lambda}^{\dot{\alpha}}
\frac{\partial}{\partial\tilde{\lambda}^{\dot{\alpha}}} -\tilde{\eta}^A
\frac{\partial}{\partial \tilde{\eta}^A} + 2\,.
\]
In the algebra $\alg{psu}(2,2|4)$, the central charge operator $\alg{C}$ is set
to zero. If one however considers the central extension to $\alg{su}(2,2|4)$,
the operator $\alg{C}$ does not need to vanish any more. 

A function $\mathcal{Y}(\{\lambda_i,\tilde{\lambda}_i,\tilde{\eta}_i\}),$
$i=1,\dots,n$ is called a Yangian invariant if it is annihilated by the
maximally iterated coproduct $\copro^{n-1}(\dots)$ for all operators $\why{J}$
and $\why{\hat{J}}$. In particular, for the charge operator the condition reads
\[
\sum_{i=1}^n \why{C}_i\,
\mathcal{Y}(\{\lambda_i,\tilde{\lambda}_i,\tilde{\eta}_i\}) = 0.
\label{eq:cc_yang}
\]
One can understand the eigenvalue of the operator $\why{C}_i$ as an additional
scaling weight for the $i$-th copy of the superspace the Yangian invariant is
defined upon. In particular the variable $\lambda_i$ now scales with a factor
$\alpha^{(1+c_i)}$. With nonzero $c_i$ one is effectively working on a weighted
projective space. 

%%%%%%%%%%%%%%%%%%%%%%%%%%%%%%%%%%%%%%%%%%%%%%%%%%%%%%%%%%%%%%%%%%%%%%%%%%%%%%%%

\subsection{Yangian invariants from undeformed building blocks}
\label{sec:undeformed}

Yangian-invariant objects for constructing scattering amplitudes in $\superN=4$
sYM theory have been explored in ref.~\cite{ArkaniHamed:2012nw} in order to derive an
all-loop generalisation of the BCFW recursion relations
\cite{Britto:2004ap,Britto:2005fq,ArkaniHamed:2008gz,Brandhuber:2008pf}.  

The main tool in the investigation of Yangian invariants
in ref.~\cite{ArkaniHamed:2012nw} are undeformed on-shell graphs, which are planar
graphs obtained by gluing two types of basic Yangian-invariant building blocks.
The building blocks used are the $\mathrm{MHV}$ and $\overline{\mathrm{MHV}}$
three-point tree-level superamplitudes in $\superN=4$ sYM theory.

After reviewing the undeformed formalism in the current subsection, we are going to
study deformed building blocks and check under which circumstances the
corresponding deformed on-shell graphs are Yangian invariants.

%%%%%%%%%%%%%%%%%%%%%%%%%%%%%%%%%%%%%%%%
\paragraph{Undeformed vertices}

Expressed in spinor-helicity and Grassmann variables of the on-shell superspace
introduced in \subsecref{YangianIntroduction}, the two basic building blocks
used in ref.~\cite{ArkaniHamed:2012nw} read
\[
\mathcal{A}_{3,\mathrm{MHV}} = \frac{\delta^4(P)\,\delta^8 (Q)}{\sprods{1}{2}
  \sprods{2}{3} \sprods{3}{1}}, \qquad \mathcal{A}_{3,\overline{\mathrm{MHV}}} =
\frac{\delta^4(P)\,\delta^4 (\tilde{Q})}{\cprods{1}{2} \cprods{2}{3}
  \cprods{3}{1}},
\label{eq:3vertices_definition}
\]
where $P^{\alpha\dot{\alpha}}:=\sum_{i=1}^3 \lambda^\alpha_i
\tilde{\lambda}^{\dot{\alpha}}_i$ is the total four-momentum, whereas
$Q^{A\alpha}:=\sum_{i=1}^3 \lambda_i^\alpha \tilde{\eta}_i^A$ and
$\tilde{Q}^A:=\cprods{1}{2} \tilde{\eta}_3^A
+\cprods{2}{3}\tilde{\eta}_1^A+\cprods{3}{1} \tilde{\eta}_2^A$. Furthermore,
$\sprods{i}{j}=\lambda_i^{\alpha}\lambda_{j\alpha}$ and
$\cprods{i}{j}=\tilde{\lambda}_{i\dot{\alpha}}\tilde{\lambda}_j^{\dot{\alpha}}$,
where indices are raised and lowered with dotted and undotted totally
antisymmetric $2\times2$ tensors $\epsilon$, where
$\epsilon_{12}=\epsilon_{\dot{1}\dot{2}}=-1$.

Being superamplitudes in $\superN=4$ sYM theory, these two three-vertices are
Yangian invariants, as shown for example in
refs.~\cite{Drummond:2009fd,Beisert:2010jq}. However, for now we would like to
focus on their symmetries without making reference to their amplitude
properties. 

We will represent $\mathcal{A}_{3,\mathrm{MHV}}$ and
$\mathcal{A}_{3,\overline{\mathrm{MHV}}}$ with a black and a white dot,
respectively,
\[
\includegraphicsboxex{vertexpair.mps}\,.
\label{fig:3verts}
\]
%

%%%%%%%%%%%%%%%%%%%%%%%%%%%%%%%%%%%%%%%%
\paragraph{Undeformed on-shell graphs}

In this paragraph we briefly review the formalism of on-shell graphs (or
diagrams). In particular we will address how to build higher-point Yangian
invariants by gluing the basic building blocks introduced in
\eqnref{eq:3vertices_definition}.

Let us consider two trivalent graphs for simplicity. If we consider their
product with one leg identified,
\[
  \mathcal{F}_5(p_1,p_2,p_3,p_4,p_I):= \mathcal{A}_{3,\mathrm{MHV}}(p_1,p_2,p_I)\,
  \mathcal{A}_{3,\overline{\mathrm{MHV}}} (-p_I,p_3,p_4),
\]
then it is not clear how to treat the state corresponding with momentum $p_I$. 
Explicitly, if we consider $I$ as an external state, then the total momentum is
$\sum_{i=1}^4\why{P}^{\alpha\dot{\alpha}}_i+\why{P}^{\alpha\dot{\alpha}}_I$, which
clearly does not annihilate $\mathcal{F}_5$. On the other hand, if we consider
the external states to be just $1,2,3$ and $4$, the total momentum is
$\sum_{i=1}^4\why{P}^{\alpha\dot{\alpha}}_i$ and it does annihilate
$\mathcal{F}_5$, but now $p_I$ acts as a ``preferred'' momentum that breaks
Lorentz invariance%
\footnote{It can be seen as follows: the spinor products
  \emph{are} Lorentz invariant (since the delta functions allow us to substitute
  $p_I=p_1+p_2$), but a Lorentz transformation acting only on $1,2,3,4$ does not
  leave the delta functions invariant.}. %
Therefore, the mere identification of a leg between two superconformal
invariants does not give another superconformal invariant.

In order to render the combination of
$\mathcal{A}_{3,\mathrm{MHV}}(p_1,p_2,p_I)$ and
$\mathcal{A}_{3,\overline{\mathrm{MHV}}} (-p_I,p_3,p_4)$ a superconformal
invariant, one has to integrate over the on-shell superspace corresponding to
the identified leg with the measure
\[
\label{eq:intmeasure}
\int \frac{\mathrm{d}^2 \lambda_I\,\mathrm{d}^2
\tilde{\lambda}_I}{\mathrm{Vol}[\grp{GL}(1)]}\,\mathrm{d}^4 \tilde{\eta}_I,
\]
a procedure which we will call \emph{gluing}.

Invariance under the total momentum generator
$\why{P}^{\alpha\dot{\alpha}}=\sum_{i=1}^4\why{P}^{\alpha\dot{\alpha}}_i$ is ensured
by the overall delta $\delta(\sum_{i=1}^4 p_i)$.  The resulting glued object can
be checked to be invariant under superconformal as well as dual superconformal
transformations provided the building blocks are Yangian invariants, as shown
in ref.~\cite{ArkaniHamed:2010kv}.

Correspondingly, one can introduce a graphical description of Yangian
invariants in terms of so-called \emph{on-shell graphs}: connected planar
graphs constructed by gluing trivalent black and white vertices associated to
$\mathcal{A}_{3,\mathrm{MHV}}$ ($\mathcal{A}_{3,\overline{\mathrm{MHV}}}$) as
in \eqnref{fig:3verts} and where each internal line corresponds to an
integration over the on-shell phase space. For example, \eqnref{eq:bwvert}
represents the graph obtained by gluing together $\mathcal{A}_{3,\mathrm{MHV}}$
and $\mathcal{A}_{3,\overline{\mathrm{MHV}}}$
\begin{equation}
  \label{eq:bwvert}
  \includegraphicsboxex{Fig_bwvert.mps} \ .
\end{equation}

\begin{figure}
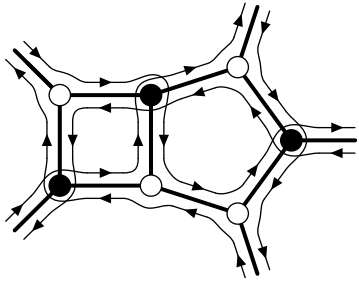

  \centering
  \includegraphicsboxex[scale=1]{Fig_5ptperm.mps}
  \caption{Permutation flow associated with a $5$-point graph.}
  \label{fig:5pts_perm}
\end{figure}

In ref.~\cite{ArkaniHamed:2012nw} it was shown how the equivalence of different
on-shell graphs can be encoded in a purely combinatorial object: the
\emph{permutation} $\sigma$ associated with the on-shell graph.  If two on-shell
graphs represent the same Yangian invariant, they will encode the same
permutation. The permutation associated with an on-shell graph is a bijective
map
\[
\label{eq:defperm}
\sigma:\{1,\dots,n\}\to\{1,\dots,n\}\ ,
\]
which is constructed as follows: starting from an external point $i$ of an
on-shell graph, follow the internal lines turning right at each black vertex and
left at each white vertex. The image $\sigma(i)$ is given by the ending external
line $j$. In \figref{fig:5pts_perm} there is an example for the permutation
obtained from following the rules described above.

\begin{figure}
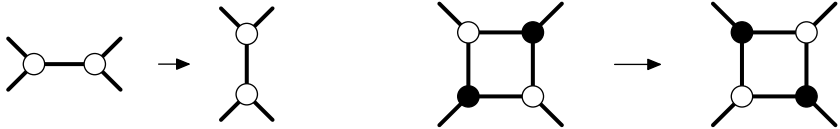

  \centering
  \begin{minipage}[h]{.4\linewidth}\centering
    \includegraphicsboxex{Figmerger.mps}
  \end{minipage}
  \begin{minipage}[h]{.4\linewidth}\centering
    \includegraphicsboxex{Figsquaremove.mps}
  \end{minipage}
  \caption{Merger and square move}
  \label{fig:moves}
\end{figure}

The equivalence of two on-shell graphs can be deduced graphically. There
are two graphical transformations that do not affect the permutation:
\emph{merging} and \emph{square move}. Merging changes the way four lines are
connected by two equally coloured vertices, while the square move rotates a
subgraph consisting of a square of vertices with alternating colours by
$90^\circ$. Both operations are depicted in \figref{fig:moves}.
%

%%%%%%%%%%%%%%%%%%%%%%%%%%%%%%%%%%%%%%%%%%%%%%%%%%%%%%%%%%%%%%%%%%%%%%%%%%%%%%%%

\subsection{Yangian invariants from deformed building blocks}
\label{sec:deformedbuildingblocks}

%%%%%%%%%%%%%%%%%%%%%%%%%%%%%%%%%%%%%%%%
\paragraph{Deformed vertices}

The undeformed building blocks $\mathcal{A}_{3,\mathrm{MHV}}$ and
$\mathcal{A}_{3,\mathrm{\overline{MHV}}}$  defined in
\eqnref{eq:3vertices_definition} do not only satisfy the central-charge
condition for Yangian invariance \eqnref{eq:cc_yang}, but the stronger
constraint
\[
\label{eq:Yangianundeformed}
\why{C}_i\cdot\mathcal{A}_{3,\mathrm{MHV}} =0,\qquad 
\why{C}_i\cdot\mathcal{A}_{3,\mathrm{\overline{MHV}}} =0,
\]
that is, the operator $\why{C}$ defined in \eqnref{eq:central_charge_op}
vanishes for each component of the tensor product individually. Taking into
account the gluing procedure above, it is not difficult to see that this
statement generalises to all Yangian invariants represented by undeformed
on-shell graphs
\[
\label{eq:undeformedC}
\why{C}_i\,\mathcal{Y} = 0
\]
for all $i$. Following the analysis of ref.~\cite{Ferro:2012xw,Ferro:2013dga}, we
will exploit the fact that the condition for Yangian invariance
\eqnref{eq:cc_yang} is less restrictive than \eqnref{eq:undeformedC}. In
particular, we will start with \emph{deformed} trivalent objects
$\mathcal{A}_\bullet$ and $\mathcal{A}_\circ$, which are not annihilated
by the central charge operators $\why{C}_i$ individually as in
\eqnref{eq:undeformedC}. After discussing under which conditions these
deformed building blocks are Yangian invariants, we will explore in which way
one can combine the deformed building blocks in order to obtain new Yangian
invariants.  

The deformation of the building blocks $\mathcal{A}_\bullet$ and
$\mathcal{A}_\circ$ will be naturally given in terms of nonzero eigenvalues
$c_i$ corresponding to each operator $\why{C}_i$. The eigenvalue $c_i$ is
referred to as the \emph{central charge} and is accompanied by an evaluation
parameter $u_i$ (see \subsecref{YangianIntroduction}) for each component $i$ of
the tensor product.  As stated at the end of \subsecref{YangianIntroduction},
in this configuration the parameters $u_k$ of the evaluation representation of
the Yangian need not be all equal.

In general, we consider the central charge to be ingoing for all external
particles. However, for each internal line one has to choose a direction for the
flow of central charge. Graphically, this is represented by an arrow, whose
reversal amounts to flipping the central charge flowing along the line while
keeping the evaluation parameter untouched:
\[
\raisebox{.4\height}{\includegraphicsboxex{FigLineConvention1.mps}}
\quad \equiv \quad
\raisebox{.4\height}{\includegraphicsboxex{FigLineConvention2.mps}} \ .
\]
It can be seen as follows: considering the line to be a 2-point invariant, the
action of the level-zero generators must satisfy
\begin{equation}
  \label{eq:2pt_l0}
  \why{J}_1 + \why{J}_2 = 0,
\end{equation}
which (for $\why{J}= \why{C}$) implies $c_1 = -c_2$. For the level one, we have
\begin{equation}
  \label{eq:2pt_l1}
  \widehat{\why{J}}_1 + \widehat{\why{J}}_2 = u_1\, \why{J}_1 + u_2 \,\why{J}_2
  = (u_1-u_2)\,\why{J}_1 = 0,
\end{equation}
meaning that $u_1=u_2$.  

For the objects $\mathcal{A}_\bullet$ and $\mathcal{A}_\circ$ we choose the
following convention:
\[
\centering
\includegraphicsboxex{vertexpair2_same.mps}
\label{fig:3verts2}
\]
which corresponds to
\[
\why{C}_i\,\mathcal{A}_\bullet = c_i \mathcal{A}_\bullet,\qquad
\why{C}_i\,\mathcal{A}_\circ = c_i \mathcal{A}_\circ.
\label{eq:C_action}
\]
The condition for Yangian invariance \eqnref{eq:cc_yang} requires the total
central charge operator to annihilate the vertex
\[
\label{eq:operatorC_action}
\sum_{i=1}^3\, \why{C}_i\cdot\mathcal{A}_\bullet =0,\qquad \sum_{i=1}^3\,
\why{C}_i\cdot\mathcal{A}_\circ =0,
\]
which translates into the condition $\zvar_1+\zvar_2+\zvar_3 = 0$ immediately.  In
\cite{Ferro:2012xw,Ferro:2013dga}, two building blocks
$\mathcal{A}_\bullet$ and $\mathcal{A}_\circ$ satisfying the condition
\eqnref{eq:C_action} have been introduced. Represented in spinor-helicity
variables, they read
\begin{align}
  \label{eq:3val_bl}
  \mathcal{A}_\bullet = \frac{\delta^4 (P)\delta^8
    (Q)}{\sprods{1}{2}^{1+\zvar_3} \sprods{2}{3}^{1+\zvar_1}
    \sprods{3}{1}^{1+\zvar_2}},\qquad
  \mathcal{A}_\circ = \frac{\delta^4 (P)
    \delta^4(\tilde{Q})}{\cprods{1}{2}^{1-\zvar_3}
    \cprods{2}{3}^{1-\zvar_1} \cprods{3}{1}^{1-\zvar_2}}\,.
\end{align}
While it is not difficult to check the invariance of $\mathcal{A}_\bullet$ and
$\mathcal{A}_\circ$ under the superconformal generators
\eqnref{eq:superconformal_onshell_superspace}, Yangian invariance is ensured
only after imposing vanishing under one level-one generator $\widehat{\why{J}}$.
We choose to consider the action of the operator
$\widehat{\why{P}}^{\alpha\dot{\alpha}}$ on
$\mathcal{A}_\bullet,\;\mathcal{A}_\circ$ in order to derive a relation between
the central charges%
\footnote{Our naming conventions differ from those used by
  the authors of ref.~\cite{Ferro:2012xw,Ferro:2013dga}; the ``spectral parameters''
of these references are called central charges here.} $\zvar_i$ and the evaluation
parameters $u_i$, which is a necessary condition for Yangian invariance.

The level-one operator $\widehat{ \why{P}}^{\alpha\dot{\alpha}}$ can be derived from
\eqnref{eq:nth_tensorproduct} and reads %\cite{Drummond:2009fd}
\begin{align}
  \label{eq:phat_def}
  \widehat{\why{P} }^{\alpha\dot{\alpha}}= \sum_{1\leq i < j \leq n}
  \Bigl[\bigl(\delta^{\dot{\alpha}}_{\dot{\beta}} \why{L}^\alpha_{i\,\beta} +
  \delta^\alpha_\beta \bar{\why{L}}^{\dot{\alpha}}_{i\,\dot{\beta}} +
  \delta^{\dot{\alpha}}_{\dot{\beta}} \delta^\alpha_\beta \why{D}_i \bigr)
  \why{P}_j^{\beta\dot{\beta}} + \bar{\why{Q}}_{i A}^{\dot{\alpha}} \why{Q}_j^{A\alpha}
  - (i\leftrightarrow j)\Bigr] + \sum_{k=1}^n u_k
  \why{P}_k^{\alpha\dot{\alpha}}\,.
\end{align}
The action of $\widehat{\why{P}}$ on $\mathcal{A}_\bullet$ is easily computed, as
the terms in square brackets in \eqnref{eq:phat_def} are all
single-derivative operators. Then, a short calculation leads to
\begin{align}
  \label{eq:Phatsd}
  \widehat{\why{P}}^{\alpha\dot{\alpha}} \cdot \mathcal{A}_\bullet &=
  \Bigl\{\bigl[ \zvar_2 + \zvar_3 + u_1 \bigr] \lambda_1^\alpha
  \tilde{\lambda}_1^{\dot\alpha} + \bigl[ \zvar_3 - \zvar_1 + u_2 \bigr]
  \lambda_2^\alpha \tilde{\lambda}_2^{\dot\alpha} + \bigl[ -\zvar_1 - \zvar_2 +
  u_3\bigr] \lambda_3^\alpha
  \tilde{\lambda}_3^{\dot\alpha}\Bigr\}\mathcal{A}_\bullet.
\end{align}
A similar equation can be derived for $\mathcal{A}_\circ$, the only difference
being the sign with which the central charges appear.  Since the $p_k$'s sum to
zero, there are just two independent equations, and the solution expressing the
$\zvar$'s in terms of the evaluation parameters reads after imposing $\zvar_1+\zvar_2+\zvar_3=0$
\begin{align}
  \label{eq:z_to_u_b}
  \mathcal{A}_\bullet:& \qquad \zvar_1 = u_2-u_3,\quad \zvar_2= u_3 -
  u_1, \quad \zvar_3 = u_1 - u_2\,; \\
  \label{eq:z_to_u_w}
  \mathcal{A}_\circ:&\qquad\zvar_1 = u_3-u_2,\quad \zvar_2= u_1 - u_3, \quad
  \zvar_3 = u_2 - u_1\,.
\end{align}
Notice that shifting all the $u$'s by a common quantity $\alpha$ has the only
effect of sending $\widehat{\why{P}} \to \widehat{\why{P}} + \alpha \,\why{P} $, and
this shift does not affect \eqnref{eq:phat_def}. Therefore, the evaluation
parameters $u_i$ are defined up to an overall shift.  

This type of analysis -- albeit in the
language of another set of variables and for a limited set of on-shell graphs
-- was first done in ref.~\cite{Ferro:2012xw,Ferro:2013dga}.

%%%%%%%%%%%%%%%%%%%%%%%%%%%%%%%%%%%%%%%%%%%%%%%%%%%%%%%%%%%%%%%%%%%%%%%%%%%%%%%%

\subsection{Deformed on-shell graphs and the permutation flow}
\label{sec:permflow}

Having introduced the deformed building blocks
$\mathcal{A}_\bullet$ and $\mathcal{A}_\circ$ in \eqnref{eq:3val_bl}, let us now
discuss how to glue them in order to obtain Yangian invariants. In parallel to
the previous discussion, a deformed on-shell graph is a connected planar graph
composed by gluing the deformed building blocks.

In contrast to the gluing procedure for the undeformed graphs built from
undeformed blocks \eqnref{eq:3vertices_definition}, we will have to take care
of the central charge $\zvar$ and the evaluation parameter $u$ associated to each
internal and external line.  Gluing two vertices with a common leg $I$ is again
done by integrating over the on-shell superspace corresponding to this
particular leg using the measure \eqnref{eq:intmeasure}.  The combined
object is a Yangian invariant with the following assignment for the internal
central charge $c_I$ and evaluation parameter $u_I$ 
\[
\label{eq:gluedobject}
\int \frac{\mathrm{d}^2 \lambda_I\,\mathrm{d}^2
  \tilde{\lambda}_I}{\mathrm{Vol}[\grp{GL}(1)]}\,\mathrm{d}^4 \tilde{\eta}_I
\mathcal{A}_\bullet\bigl(\lambda_I,\tilde{\lambda}_I,\tilde{\eta}_I,\zvar_I,u_I\bigr)\,
\mathcal{A}_\circ\bigl(\lambda_I,-\tilde{\lambda}_I,-\tilde{\eta}_I,-\zvar_I,u_I\bigr),
\]
where we omitted the dependence on the on-shell superspace variables of
particles $1,2,3,4$ in the integrand. In~\appref{app:yangian_inv} it is shown
that the identifications of $c_I$ and $u_I$ in \eqnref{eq:gluedobject}
ensure the Yangian invariance of the resulting object.

An object represented by a deformed on-shell graph is a Yangian invariant if the
central charges and evaluation parameters on each internal line are identified
as described above and eqs.~\eqref{eq:z_to_u_b},~\eqref{eq:z_to_u_w} are
satisfied at each vertex.  Imposing all those constraints simultaneously leads
to a system of linear equations relating the central charges $c_i$ and
evaluations parameters $u_i$ of the external legs.

%%%%%%%%%%%%%%%%%%%%%%%%%%%%%%%%%%%%%%%%
\paragraph{Two equally coloured vertices} 

Let us see how this works for an easy
example: imposing all constraints for the graph
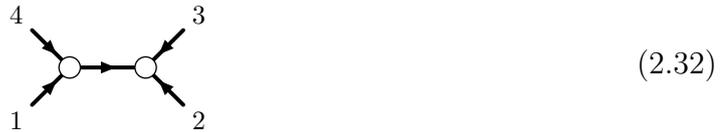
\[
\includegraphicsboxex{mergergraph1.mps}
\label{fig:mergergraph1}
\]
leads to
\begin{align}
  0&=\zvar_1+\zvar_2+\zvar_3+\zvar_4,\nln 
u_2&=\zvar_1+\zvar_2+u_1,\nln
u_3&=\zvar_1+2\zvar_2+\zvar_3+u_1,\nln 
u_4&=\zvar_2+\zvar_3+u_1.
\end{align}
It is not difficult to see that the solution to the above system could have been
obtained from another set of linear equations, corresponding to the following
graph:
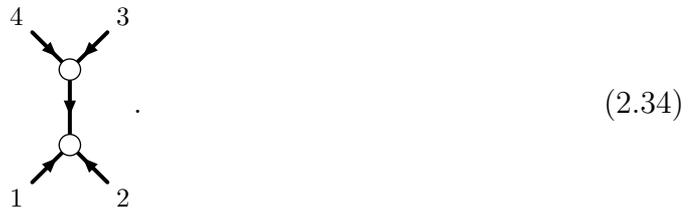
\[
\includegraphicsboxex{mergergraph2.mps}.
\label{fig:mergergraph2}
\]
This does not come unexpectedly: the merger operation is valid for deformed
on-shell graphs as well, as already shown in ref.~\cite{Ferro:2012xw,Ferro:2013dga}.

%%%%%%%%%%%%%%%%%%%%%%%%%%%%%%%%%%%%%%%%
\paragraph{Chain of vertices}

Let us now look at another trivial configuration: a tree on-shell graph with $n$
external legs, $n_\mathrm{V}=n-2$ vertices and $n_{\mathrm{I}}=n-3$ internal lines, for
example
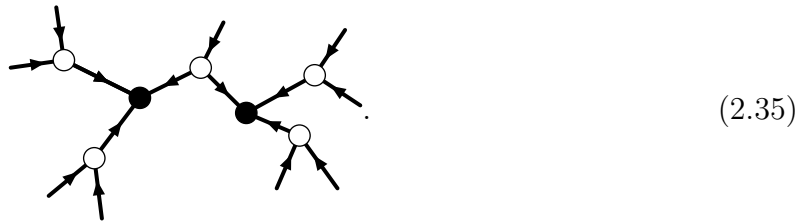
\[
\includegraphicsboxex{treegraph.mps}.
\label{fig:treegraph}
\]
As discussed in \subsecref{sec:deformedbuildingblocks}, there are three free
parameters associated to each vertex. This turns the consideration of a tree
graph into an easy problem: taking the two conditions for gluing an internal
line into account a simple counting shows that the number of independent
quantities for the Yangian invariant in \eqnref{fig:treegraph} equals the
number $n$ of external lines:
\begin{equation}
  n=  3\,n_\mathrm{V}-2\,n_\mathrm{I}\,.
\end{equation}

%%%%%%%%%%%%%%%%%%%%%%%%%%%%%%%%%%%%%%%%
\paragraph{Four-point graph, four vertices} 

\begin{figure}
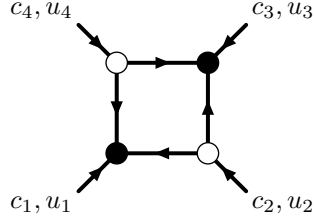

  \centering
  \includegraphicsboxex{Fig4pt.mps}
  \caption{Four-point on-shell graph}
  \label{fig:4pt}
\end{figure}

In order to produce more interesting configurations, one will need to build
on-shell diagrams containing loops. Let us have a look at the simplest case: a
box with alternating white and black dots depicted in \figref{fig:4pt}.

Solving the linear system corresponding to the four-point on-shell graph in
\figref{fig:4pt} leads to
\[
c_1=-c_3,\quad c_2=-c_4\quad\text{and}\quad u_1=u_3,\quad u_2=u_4\,.
\]
Thus the values of the evaluation parameters and central charges can be
identified between legs 1 and 3 as well as legs 2 and 4 respectively. Note
furthermore that the condition in \eqnref{eq:cc_yang} is trivially satisfied
with the above solution.

\begin{figure}
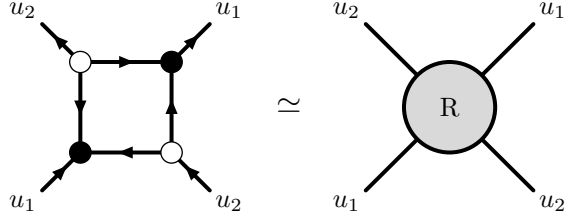

  \centering
  \includegraphicsboxex{Fig4pt_rma.mps}$\quad\simeq\quad$\includegraphicsboxex{Fig4pt_rmb.mps}
  \caption{The four-point tree-level deformed on-shell graph has a structure
    resembling the R-matrix $\mathrm{R}(u_1-u_2)$ of quantum integrable
    systems. Note that in order to match the usual conventions for R-matrices,
    we have reversed the arrows of the four-point on-shell diagram with respect
    to \protect\figref{fig:4pt}. }
  \label{fig:4pt_rm}
\end{figure}

The construction and study of this four-point deformed on-shell graph was
performed in ref.\cite{Ferro:2012xw,Ferro:2013dga}, where it was argued that it
intertwines the external states. In quantum integrable systems, this is the
r\^ole played by the R-matrix, see \figref{fig:4pt_rm}.  It depends on the
difference of the evaluation parameters of the external states, which is a
natural fact in the context of quantum integrable systems.

Let us then explore the properties of R-matrices in quantum integrable systems
a little closer before translating their properties back to the language of
on-shell diagrams. Note that all the following equations featuring on-shell
graphs are to be understood as ``the linear system for the graph on the
right-hand side is equivalent to the linear system represented on the left-hand
side''.
\begin{itemize}
\item R-matrices satisfy the Yang--Baxter equation:
    \[
    \mathrm{R}_{12}(u_1-u_2)\mathrm{R}_{13}(u_1-u_3)\mathrm{R}_{23}(u_2-u_3)=
    \mathrm{R}_{23}(u_2-u_3)\mathrm{R}_{13}(u_1-u_3)\mathrm{R}_{12}(u_1-u_2)
    \]
or in terms of R-matrix and on-shell diagrams:
      \[
      \includegraphicsboxex[scale=1.2]{RMatrix1a.mps}=\includegraphicsboxex[scale=1.2]{RMatrix1b.mps}
      \hspace{2cm}
      \includegraphicsboxex[scale=1.2]{RMatrix4b.mps}=\includegraphicsboxex[scale=1.2]{RMatrix4a.mps}.
      \]
 \item Iterated application is proportional to the identity
    \[
    \mathrm{R}_{12}(u_1-u_2)\mathrm{R}_{21}(u_2-u_1)=f(u_1,u_2)\cdot I
\hspace{1.5cm}
      \includegraphicsboxex[scale=1.2]{RMatrix2a.mps}\sim\includegraphicsboxex[scale=1.2]{RMatrix2b.mps}
	\hspace{1cm}
      \includegraphicsboxex[scale=1.2]{RMatrix5a.mps}\sim\includegraphicsboxex[scale=1.2]{RMatrix5b.mps},
      \]
    where the function $f(u_1,u_2)$ is some function of the two spectral parameters. 
\end{itemize}

\begin{figure}
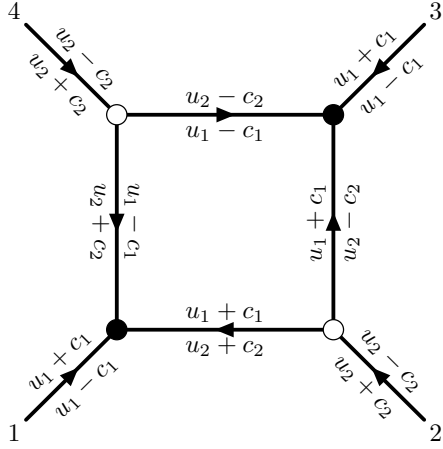

  \centering
  \includegraphicsboxex[scale=.95]{Fig4ptupm_v3.mps}
  \caption{Four-point on-shell graph with flows of $u^\pm$ indicated}
  \label{fig:4ptupm}
\end{figure}

\begin{figure}
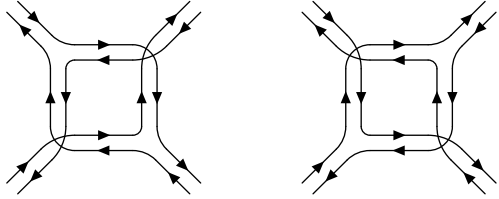

  \centering
  \includegraphicsboxex{DoubleLineFig4pt.mps}
  \hspace{1cm}
  \includegraphicsboxex{DoubleLineFig4ptsquaremove.mps}
  \caption{Four-point on-shell graph in the double-line formalism and its
    square-moved version. The square simply switches the r\^ole of the flow: a
    line going around the loop clockwise will do so counterclockwise after the
    square move.}
  \label{fig:4ptdl}
\end{figure}

After identifying the four-point Yangian invariant in~\figref{fig:4pt} with 
%the $\mathrm{R}$-matrix 
$\mathrm{R}({u_1-u_2})$, it is worthwhile to
investigate the diagram further. In particular, one can define new variables
\[
  \label{eq:upm}
u^\pm=u\pm c\,,
\]
which are convenient for tracking how central charges $c_i$ and evaluation
parameters $u_i$ flow through on-shell diagrams. The quantity $u^+$ will be
written to the left of a line looking in the direction of its arrow, while $u^-$
will be placed to the right. This convention is compatible with flipping the
direction of the arrow using the rules in \subsecref{sec:deformedbuildingblocks}
\[
  \vspace{.1cm}
  \includegraphicsboxex{upmConventions.mps}\quad.
\]
After solving the linear system one finds the configuration in~\figref{fig:4ptupm}.
Tracing the quantities $u^\pm$ through the diagram and connecting them by
lines, one can easily recognise the following rules for traversing: 
\begin{itemize}
  \item at a black vertex, turn right,
  \item at a white vertex, turn left. 
\end{itemize}
Thus it is suggestive to keep track of
this information in terms of the following double-line formalism. Hereby the
black and white vertices translate into
\[
  \includegraphicsboxex{DoubleLineBlack.mps}
  \hspace{2cm}
  \includegraphicsboxex{DoubleLineWhite_v2.mps}
\]
and \figref{fig:4ptupm} becomes \figref{fig:4ptdl}, where we did not draw the
vertices and their connecting lines for convenience.  Comparing the above rules
with the definition of the permutation represented by an on-shell graph
introduced in \subsecref{sec:undeformed}, it becomes obvious that the quantities
$u^\pm$ follow exactly the lines determining the permutations.  In other words,
the permutation map $\sigma$ defined in \eqnref{eq:defperm} above keeps track
of the flow of the $u^\pm$ between external legs.  The two vertices defined
above are sufficient to translate any on-shell graph into the linear system of
equations ensuring Yangian invariance immediately, as will be shown below.

The four-point diagram discussed here is the subgraph appearing in the
square-move transformation of the undeformed on-shell graph depicted in
\figref{fig:moves}. Employing the double-line formalism, it is trivial to see
that his transformation does not alter the permutation in the deformed setup
either: switching the r\^ole of black and white vertices amounts to switching
the path a quantity takes around the loop (see \figref{fig:4ptdl}), while the
solution to the conditions ensuring Yangian invariance remains unchanged.

%%%%%%%%%%%%%%%%%%%%%%%%%%%%%%%%%%%%%%%%
\paragraph{Five points and seven vertices} 

In order to check the identification between the flow of $u^\pm$ and the
permutation encoded in an on-shell graph, let us consider another example: the
five-point on-shell diagram 
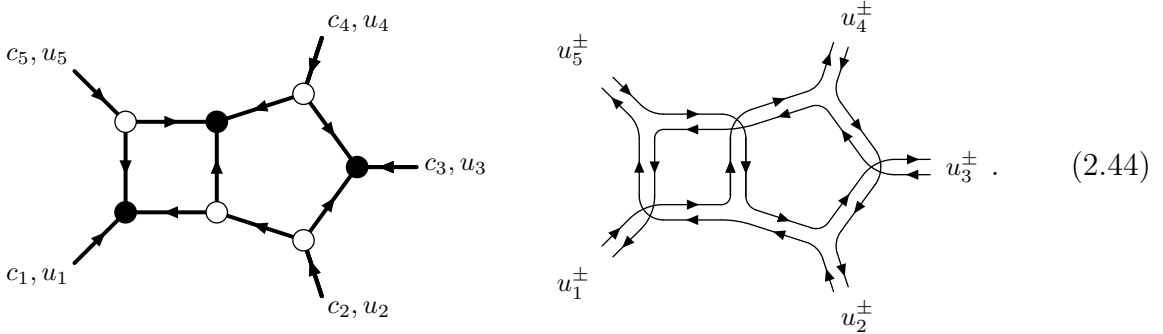
\[
  \includegraphicsboxex{Fig5pt.mps}\,
  \qquad
  \includegraphicsboxex{DoubleLineFig5pt.mps} \ .
  \label{fig:5pt}
\]
Imposing Yangian invariance by solving the linear system yields
\[
  \{u^+_1,u^+_2,u^+_3,u^+_4,u^+_5\}=\{u^-_4,u^-_5,u^-_1,u^-_2,u^-_3\}\,,
\]
which corresponds to the permutation $(1,2,3,4,5)\rightarrow(4,5,1,2,3)$ -- a
simple cyclic shift by three.  

The same result is implied by the second diagram in the above picture.
With the double-line formalism it is obvious that the linear system obtained by
imposing Yangian invariance of the corresponding on-shell diagram will fix at
most half of the $2n$ free variables $c_i$ and $u_i$, $i=1\ldots n$. The
solution to the linear system can be expressed in the permutation $\sigma$
defined in \subsecref{sec:undeformed}. Considering all external particles as
ingoing, the solution simply reads
\[
 u^-_{\sigma(i)} = u^+_i\,.
  \label{eq:solution}
\]

%%%%%%%%%%%%%%%%%%%%%%%%%%%%%%%%%%%%%%%%%%%%%%%%%%%%%%%%%%%%%%%%%%%%%%%%%%%%%%%%
\subsection{Superconformal and Yangian Anomalies}
\label{app:asdasd}

In this subsection we would like to comment on the issue of exactness of
Yangian invariance and on the uniqueness of the invariants we have constructed.

A common belief in integrable systems is that the extended symmetries present
in such systems determine the observables uniquely.  In particular, the
S-matrix should be determined uniquely by its invariance.  

In the previous parts of this section we have constructed a large collection of
Yangian invariants which are -- most likely -- almost all inequivalent.  Here
it makes sense to compare only those functions with the same external data
(number of legs, MHV degree, central charges and evaluation parameters).

It is quite clear that for a given set of external data there exist several
on-shell graphs: At a low number of loops in the on-shell diagram, 
there will only be few, if any, graphs whose permutation structure matches the external data.  
However, at higher on-shell loops, there must be further graphs 
with the same permutation structure 
because the set of permutation structures for a given number of legs is finite.
It is unlikely that all of these graphs reproduce the same function, in particular as
they are not all connected by the permissible merger and square moves (see
\figref{fig:moves}).

This argument seems to show that there are many inequivalent Yangian invariants
for the same external data. However, following the analysis in
ref.~\cite{Bargheer:2009qu} (see ref.~\cite{Bargheer:2011mm} for a review), the
flaw of the argument is that the derived functions are not actually Yangian
\emph{invariant}.  The reason is that the fundamental three-point functions
spoil superconformal and Yangian symmetry in a subtle way.  They are invariant
for almost all external \emph{momentum} configurations, but invariance breaks
down when all three momenta are exactly collinear.%
\footnote{The discussion requires a real signature of spacetime. It is most straight-forward in
$(2,2)$ split signature where three-point function actually exist, therefore
let us assume this signature.} This can be observed most easily in the
following form of the three-point vertex~\eqref{eq:3val_bl}
\begin{align}
A_\bullet &=
\operatorname{sign}\bigbrk{\sprods{1}{2}}
\int \frac{d\tilde z_1}{\tilde z_1^{1+c_1}}\,\frac{d\tilde z_2}{\tilde z_2^{1+c_2}}\,
\delta^2(\tilde z_1\lambda_1+\tilde z_2\lambda_2+\lambda_3)
\nonumber\\&\qquad\cdot
\delta^2(\tilde\lambda_1-\tilde z_1\tilde\lambda_3)\,
\delta^2(\tilde\lambda_2-\tilde z_2\tilde\lambda_3)\,
\delta^4(\tilde\eta_1-\tilde z_1\tilde\eta_3)\,
\delta^4(\tilde\eta_2-\tilde z_2\tilde\eta_3).
\end{align}
The sign factor multiplying the integral is responsible
for conformal ``anomaly'' at collinear momenta
$\delta A_\bullet=\delta \operatorname{sign}\bigbrk{\sprods{1}{2}}\cdot\ldots\,$.
In particular, the deformations in the weights 
w.r.t.\ $\tilde z_1$ and $\tilde z_2$ do not interfere
with the symmetry breaking.

The anomaly of an on-shell graph is the sum of the anomalies 
of all constituent vertices.%
\footnote{This is the behaviour for $(2,2)$ split signature.
For $(3,1)$ Minkowski signature the anomaly works differently,
and requires a suitable combination of contributions to the S-matrix,
see ref.~\cite{Bargheer:2012cp} for further comments.}
\[
\delta
\Biggl(\,\includegraphicsboxex{App1.mps}\Biggr)
= \sum_\text{vertices}\,
\includegraphicsboxex{App2.mps} \ .
\]
This form demonstrates that the overall anomaly depends on the precise
structure of the on-shell graph.  Two inequivalent graphs with equal external
structure have a different anomaly structure, and therefore they are expected
to have a different functional dependence. This is most evident if the graphs
have a different number of on-shell loops.  Moreover one may wonder whether the
merger and square moves are in fact \emph{exact} equivalences of graphs in the
above sense. 

%%%%%%%%%%%%%%%%%%%%%%%%%%%%%%%%%%%%%%%%%%%%%%%%%%%%%%%%%%%%%%%%%%%%%%%%%%%%%%%%
%%%%%%%%%%%%%%%%%%%%%%%%%%%%%%%%%%%%%%%%%%%%%%%%%%%%%%%%%%%%%%%%%%%%%%%%%%%%%%%%

\section{On-shell graphs and scattering amplitudes}
\label{sec:amplitudes}

In the preceding section we developed the tools to construct deformed Yangian
invariants. In the following section, we want to investigate whether it is
possible and meaningful to introduce deformed scattering amplitudes as (sums
of) deformed on-shell graphs preserving manifest Yangian invariance.

We start with a short review of the formalism linking undeformed on-shell
graphs with ordinary scattering amplitudes; the interested reader can find the complete
formalism in ref.~\cite{ArkaniHamed:2012nw} and references therein. 
We then investigate the possibility to deform tree-level amplitudes 
using deformed Yangian-invariant on-shell graphs.
Finally, following the spirit of
\cite{Ferro:2012xw,Ferro:2013dga}, we investigate whether it is possible to
use the deformation parameters as Yangian-preserving regulators 
for the four-point one-loop amplitude of $\superN=4$ sYM\@.
We discuss our findings at the end of the section.

%%%%%%%%%%%%%%%%%%%%%%%%%%%%%%%%%%%%%%%%%%%%%%%%%%%%%%%%%%%%%%%%%%%%%%%%%%%%%%%%

\subsection{Grassmannian formalism}
\label{sec:grassm}

In this subsection we briefly review the Grassmannian formulation of scattering
amplitudes in $\superN=4$ sYM -- first introduced in ref.~\cite{ArkaniHamed:2009dn}
-- and its relation to on-shell graphs, which has been developed in
ref.~\cite{ArkaniHamed:2012nw}. The discussion assumes familiarity with the BCFW
recursion relations and its supersymmetric extension
\cite{Britto:2004ap,Britto:2005fq,ArkaniHamed:2008gz,Brandhuber:2008pf} as well
as the concepts of helicity amplitudes.

%%%%%%%%%%%%%%%%%%%%%%%%%%%%%%%%%%%%%%%%

\paragraph{Grassmannian manifolds}

The \emph{Grassmannian manifold} $G(k,n)$ (from now on simply ``Grassmannian'')
is the space of complex $k$-planes in $\mathbb{C}^n$ passing through the origin.
It is a generalisation of the notion of complex projective space; in particular,
$\mathbb{CP}^{n-1}$ is the Grassmannian manifold $G(1,n)$. A $k$-plane in
$\mathbb{C}^n$ is uniquely determined by a set of $k$ linearly independent
complex $n$-tuples, therefore one can identify a point in $G(k,n)$ with an
equivalence class of $k\times n$ complex matrices $C$ of rank $k$. In fact,
given such a matrix $C$, the right action of $\grp{GL}(k)$ maps $C$ to a
different matrix $C'$. Since $C$ and $C'$ identify the same $k$-plane in
$\mathbb{C}^n$, we can identify the Grassmannian $G(k,n)$ with the space of
$k\times n$ rank-$k$ complex matrices modulo a $\grp{GL}(k)$ rescaling. The
matrix represents the $k$ $n$-vectors that identify the plane in $\mathbb{C}^n$.

Let us introduce two further notions used in the Grassmannian language: a
\emph{top-cell} of the above chart of the Grassmannian $G(k,n)$ is the cell
where all sets of $k$ consecutive columns of the matrix $C$ become linearly
independent. Its dimension is the full dimension $k(n-k)$ of $G(k,n)$. On the
contrary, a \emph{generic} cell is allowed to have sets of linearly dependent
consecutive columns in its matrix $C$.  The \emph{stratification} of the
Grassmannian is then defined to be the system of nested boundaries of cells,
where a boundary is described by the above linearly dependent configuration of
columns. The decomposition of cells depends on the choice of coordinates on the
Grassmannian $G(k,n)$.

%%%%%%%%%%%%%%%%%%%%%%%%%%%%%%%%%%%%%%%%

\paragraph{Scattering amplitudes as integrals over Grassmannians}

The authors of refs.~\cite{ArkaniHamed:2009dn,ArkaniHamed:2012nw} showed that
it is possible to express tree-level as well as loop amplitudes in $\superN=4$
sYM as integrals over Grassmannian manifolds. In particular, the tree-level
amplitude can be written as the following integral
\[
  \label{eq:grassm_tree}
  \mathcal{A}_{k,n} \,:=\,
  \int \frac{\mathrm{d}^{k\times n}C_{r a}}{\mathrm{Vol}[\grp{GL}(k)]}
  \frac{1}{(12\dots k) (23\dots k+1)\dots (n-1\dots k-1)} \,
  \prod_{r=1}^k \delta^{4|4} \Bigl(\sum_{a=1}^n C_{r a}\,\mathcal{Z}_a\Bigr),
\]
where $(i\dots i+k)$ is the $i$-th $k$-minor of the matrix $C$ and $\mathcal{Z}
= (\lambda,\mu;\tilde{\eta})$ are supertwistor variables, as introduced in
\cite{Witten:2003nn}.  Most of the integrations in \eqnref{eq:grassm_tree} are
fixed by the bosonic delta functions; the number of bosonic integrals left to be
evaluated via contour integration is $(k-2) (n-k-2)$.  Notice that
\eqnref{eq:grassm_tree} makes superconformal symmetry manifest, since supertwistor
variables $\mathcal{Z}$ transform linearly under $\grp{PSU}(2,2|4)$.

In order to proceed to loop amplitudes it is useful to introduce yet another
way of expressing on-shell graphs. In
ref.~\cite{ArkaniHamed:2012nw} it was shown that any on-shell graph with
$n_{\mathrm{F}}$ faces can be associated with an integral of an
$(n_{\mathrm{F}}-1)$-form defined on the Grassmannian $G(k,n)$, where the
parameters $k$ and $n$ are related to the on-shell graph via
\[
  k = n_\mathrm{w} + 2\, n_\mathrm{b} - n_\mathrm{i}\qquad\text{and}\qquad n = 3\,(n_\mathrm{b}+n_\mathrm{w}) - 2 \, n_\mathrm{i} \ .
\]
Here, $n_\mathrm{w}$ denotes the number of white vertices, $n_\mathrm{b}$ the
number of black vertices and $n_\mathrm{i}$ the number of internal lines.

Each on-shell graph can be associated with a matrix $C$ denoting a cell of the
Grassmannian, which in turn is a function of $n_{\mathrm{F}}-1$ auxiliary
\emph{face variables} $f_i$. Expressed in these face variables, the integral
associated with the on-shell graph is obtained by integrating over all internal
lines and takes the form
\[\label{eq:facemeas}
  \prod\frac{\mathrm{d}f_i}{f_i}\,.
\]
Accordingly, amplitudes are constructed by supplementing
the measure \eqnref{eq:facemeas} with the appropriate delta functions coming from
the gluing of three-vertices. This leads to the integral
\[
  \label{eq:onshell_int}
  \mathcal{I}_{\mathrm{graph}}\,=\, \int \biggl[\prod_{i=1}^{n_{\mathrm{F}}-1}\frac{\mathrm{d}f_i}{f_i}\biggr]
  \, \delta^{2k} \Bigl( \sum_{a=1}^n C_{r a}(f_i) \tilde{\lambda}_a \Bigr)\,
  \delta^{2(n-k)}\Bigl( \sum_{s=1}^{n-k} C_{s a}(f_i) \lambda_s \Bigr)\,
  \delta^{0|4k}  \Bigl( \sum_{a=1}^n C_{r a}(f_i) \tilde{\eta}_a \Bigr).
\]
The above integral is a generalisation of \eqnref{eq:grassm_tree}, which in
addition allows to describe loop amplitudes.  The precise connection between
undeformed on-shell graphs and (loop) amplitudes is nontrivial.  We do not
discuss the relation between face variables and the matrix $C$
representing a point in the Grassmannian here. The interested reader may
consult ref.~\cite{ArkaniHamed:2012nw} for the complete formalism, or
ref.~\cite{Ferro:2012xw,Ferro:2013dga} for a concise review in the light of the
deformations introduced in the following sections.

%%%%%%%%%%%%%%%%%%%%%%%%%%%%%%%%%%%%%%%%

\paragraph{All-loop BCFW recursion relations in the Grassmannian formalism} 

In ref.~\cite{ArkaniHamed:2012nw} it was pointed out that the BCFW recursion
relations in the Grassmannian formalism are obtained as solutions to a formal
boundary equation. This equation states that the singularities of an amplitude
in $\superN=4$ sYM theory are given solely by factorisation channels and
forward limits, where
\begin{itemize}
\item a factorisation channel refers to the splitting of an amplitude in the
  special kinematical situation where the sum of consecutive momenta becomes
  on-shell;
\item a forward limit of an on-shell $l$-loop amplitude is an $(l-1)$-loop
  amplitude with two additional legs with opposite momentum. We refer to
  ref.~\cite{CaronHuot:2010zt} for the analysis of the r\^{o}le played by
  forward limits in the description of loop amplitudes.
\end{itemize}
The solution to this formal boundary equation is a sum of on-shell diagrams,
each of which corresponds to a BCFW channel. In particular, the solution
provides the integrand for the associated scattering amplitude\footnote{Upon
  dimensional regularisation, additional terms breaking dual conformal
  invariance can appear \cite{Bern:2008ap}. These terms are not present in the
  purely four-dimensional analysis such as the one-loop amplitude considered
  below. }.  Since each summand is a Yangian invariant, this proves that the
integrand for any loop amplitude in $\superN=4$ sYM theory is a Yangian
invariant modulo partial integration~\cite{Drummond:2010qh,Drummond:2010uq}.
However, loop amplitudes in $\superN=4$ sYM theory do not exhibit Yangian
invariance, as dealing with IR divergences breaks conformal invariance.

Conveniently, the on-shell graphs representing the BCFW channels for a
tree-level amplitude $\mathcal{A}_{n,k}$ can as well be expressed as boundaries
of the top-cell%
\footnote{%
  It is possible to show (see ref.~\cite{Postnikov:math0609764}) that an
  on-shell graph representing the top-cell for a certain $k$ and $n$ corresponds
  to a permutation which is just a cyclic shift by $k$. Furthermore, in the
  aforementioned paper it was explained how to construct a representative
  on-shell graph for the top-cell.} %
of a certain codimension. The boundaries of a cell are obtained
graphically by removing specific edges (called \emph{removable edges} in
ref.~\cite{ArkaniHamed:2012nw}) from the associated on-shell graph. 

The identification of the boundaries of the cell associated with a given
on-shell graph together with the BCFW construction allows to express any
tree-level amplitude starting from the top-cell.  

For loop amplitudes there is no analogue of a top-cell object, that is, an
on-shell graph whose boundaries directly yield the BCFW channels.

%%%%%%%%%%%%%%%%%%%%%%%%%%%%%%%%%%%%%%%%%%%%%%%%%%%%%%%%%%%%%%%%%%%%%%%%%%%%%%%%

\subsection{Top-cell vs.\ BCFW decomposition, deformation and compatibility}
\label{sec:topcell}

In the following subsection we will investigate the possibility of linking
deformed on-shell graphs with ``deformed'' amplitudes in $\superN=4$ sYM\@. 
In order to do so, we will first review the tools developed in
ref.~\cite{Ferro:2012xw,Ferro:2013dga} that relate a Grassmannian integral to a
deformed on-shell graph. Subsequently we will turn to the interplay between
on-shell graphs, BCFW decomposition and Yangian invariance.

%%%%%%%%%%%%%%%%%%%%%%%%%%%%%%%%%%%%%%%%

\paragraph{Grassmannian integral for deformed scattering amplitudes}

\begin{figure}
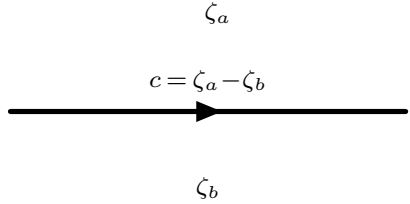

  \centering
  \includegraphicsboxex[scale=1.3]{Fig_facevars.mps}
  \caption{Face variables and central charges}
  \label{fig:facevar_def}
\end{figure}

The integral corresponding to a deformed on-shell graph is obtained by deforming
the ordinary measure \eqnref{eq:facemeas}. In terms of the face variables
$f_i$, the deformation reads
\[
  \label{eq:faceshift}
  \prod_{i=1}^{n_{\mathrm{F}}} \frac{\mathrm{d}f_i}{f_i}\;\to\; \prod_{i=1}^{n_{\mathrm{F}}}
  \frac{\mathrm{d}f_i}{f_i^{1+\zeta_i}}\,.
\]
Here the \emph{face shifts} $\zeta$ are dual variables with respect to the
central charge $c$; the latter is obtained from the difference of the two
adjacent face shifts%
\footnote{For a three-point vertex this identity makes the $\zvar$'s have the
  same relationship to the $u$'s as to the $\zeta$'s.  Therefore, the $\zeta$'s
  must agree locally with the $u$'s up to an overall shift.  Note that the
  latter shift depends on the position and hence the $\zeta$'s are not
equivalent to the $u$'s.}
\[
\label{eq:faceshift2}
\zeta_a - \zeta_b = c
\]
as depicted in \figref{fig:facevar_def}. It is evident that the resulting deformed
integrand is not a meromorphic function any more.  This is an important fact,
since it questions a clear interpretation of the BCFW decomposition of the
deformed amplitudes.

\paragraph{Deformed scattering amplitudes}

In the context of $\superN=4$ sYM theory, it makes sense to try to construct
Yangian-invariant deformed scattering amplitudes. Naturally, those deformed
amplitudes should not violate unitarity, that is, they should have the correct
singularity structures. In the undeformed case, this is ensured via the BCFW
construction. As pointed out in the previous subsection, the BCFW recursion
relations simultaneously maintain Yangian invariance.

In order to define a deformed Yangian-invariant amplitude there are two natural
approaches:
\begin{itemize}
\item One could start from the collection of on-shell diagrams corresponding to
  the undeformed BCFW channels. One would need to deform those in a
  Yangian-invariant way individually and subsequently demand compatibility
  between the deformations, meaning that the parameters $u_i$ and $c_i$
  attached to the external legs must be equal for all diagrams. The
  compatibility among the diagrams will lead to constrained
  configurations. 

\item One could deform the on-shell graphs corresponding to the
  top-cell. However, we have argued above that the integrand resulting from a
  deformed on-shell diagram is not meromorphic. 
  In terms of the Grassmannian integral 
  a boundary is equivalent to a  
  pole given by the vanishing of a minor. 
  Thus the nonmeromorphicity 
  of the integrand associated with a deformed top-cell prevents a direct BCFW-like
  decomposition. 
  It is however possible to constrain the deformation parameters in such a way
  that the removal of the edges corresponding to BCFW channels amounts to a
  simple residue integral. 
\end{itemize}

For undeformed tree-level MHV amplitudes the BCFW decomposition of the
amplitude consists of a single term and thus a single on-shell graph, which is
the top-cell. The deformation of this on-shell graph directly translates into
the deformation of the associated amplitude. For example, the on-shell graph
depicted in \figref{fig:4ptupm} corresponds to the deformed four-point
tree-level amplitude. 

For general tree-level amplitudes, these two procedures could in principle lead
to different deformations of the amplitude, since in the second case it is not
necessarily true that the resulting channels are Yangian invariant by
themselves. Moreover, it could (and will) also happen that the only possible
solution is the undeformed BCFW decomposition. In the following subsection we
will study an easy example and analyse the results originating in the two
approaches.

Considering general loop amplitudes, the definition of a top-cell--like object
is not clear.  Only in the case of the one-loop four-point amplitude there seems
to be the analogue of a top-cell. We will investigate Yangian invariant
deformations of this particular on-shell graph in \subsecref{sec:oneloop} below.

%%%%%%%%%%%%%%%%%%%%%%%%%%%%%%%%%%%%%%%%%%%%%%%%%%%%%%%%%%%%%%%%%%%%%%%%%%%%%%%%

\subsection{Deformation of the six-point NMHV amplitude}
\label{sec:6nmhv}

\begin{figure}
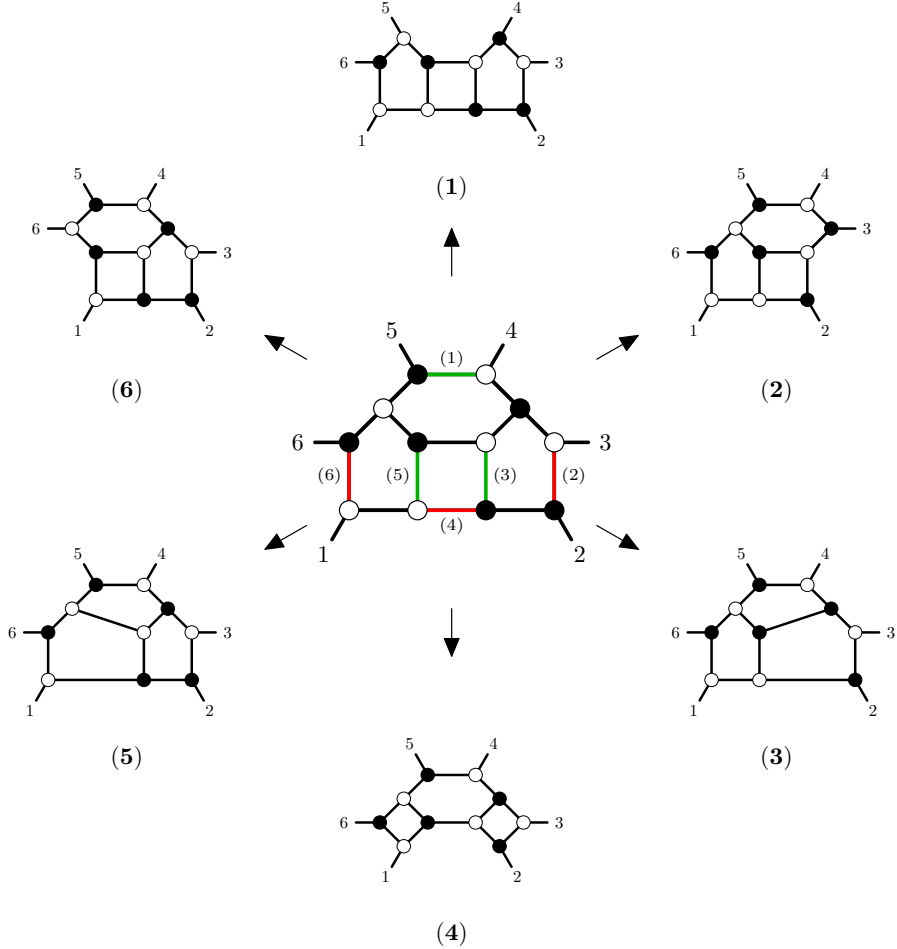

  \centering
  \includegraphicsboxex[scale=.9]{Fig6ptnmhv_bcfw.mps}
  \caption{Codimension-one boundaries of the top-cell of $G(3,6)$; the removable
    edges are highlighted.}
  \label{fig:6pttopcell_bdaries}
\end{figure}

Let us study the simplest nontrivial example: the six-point
NMHV amplitude $\mathcal{A}_{6,3}$. The simplicity of this amplitude originates
from the fact that the on-shell graphs representing the BCFW channels are
codimension-one boundaries of the top-cell (as opposed to boundaries of higher
codimension for other amplitudes). The top-cell graph associated with $G(3,6)$
and its six codimension-one boundaries are depicted in
\figref{fig:6pttopcell_bdaries}. The two possible BCFW decompositions of
$\mathcal{A}_{6,3}$ are given by adding either the contributions from graphs
$1$, $3$ and $5$ or $2$, $4$ and $6$. In the following we will choose the latter
one.

We will now investigate how to deform the on-shell graphs corresponding to
$\mathcal{A}_{6,3}$ following the two different approaches described at the end
of \subsecref{sec:topcell}.

%%%%%%%%%%%%%%%%%%%%%%%%%%%%%%%%%%%%%%%%
\paragraph{Deformation of the BCFW terms}

We will follow the first approach and impose Yangian invariance on the deformed
on-shell graphs $2$, $4$ and $6$ of \figref{fig:6pttopcell_bdaries} separately,
which amounts to satisfying the conditions implied by the double-line formalism
introduced in \subsecref{sec:permflow}.

For our choice of graphs, the permutations and their corresponding deformations
are listed in \figref{fig:bcfw}.
\begin{figure}
  \centering
  \begin{equation*}
    \begin{array}[ht]{c|c|c}
      \text{\textrm{graph}} & \text{\textrm{permutation}} & u\pm \zvar\\
      \hline &&\\
      \includegraphicsboxex[scale=.6]{Fig6pt_bcfwe.mps} &
      \begin{smallmatrix}1&2&3&4&5&6\\\downarrow&\downarrow&\downarrow&\downarrow&\downarrow&\downarrow\\4&5&6&2&1&3\end{smallmatrix}&
      \parbox{7cm}{\normalfont\footnotesize
        \begin{align*}
          u^+_1 &= u^-_4, &  u^+_2 & = u^-_5, & u^+_3 & = u^-_6,\\
          u^+_4 & = u^-_2, & u^+_5 &= u^-_1, & u^+_6 & = u^-_3,
        \end{align*}
      }\\
      &&\\
      \includegraphicsboxex[scale=.6]{Fig6pt_bcfwc.mps} &
      \begin{smallmatrix}1&2&3&4&5&6\\\downarrow&\downarrow&\downarrow&\downarrow&\downarrow&\downarrow\\3&5&6&1&2&4\end{smallmatrix}  &
      \parbox{7cm}{\normalfont\footnotesize
        \begin{align*}
          u^+_1 &= u^-_3, & u^+_2 & = u^-_5, & u^+_3 & = u^-_6, \\
          u^+_4 &= u^-_1, & u^+_5 & = u^-_2, & u^+_6 & = u^-_4,
        \end{align*}
      }
      \\
      &&\\
      \includegraphicsboxex[scale=.6]{Fig6pt_bcfwb.mps} &
      \begin{smallmatrix}1&2&3&4&5&6\\\downarrow&\downarrow&\downarrow&\downarrow&\downarrow&\downarrow\\4&6&5&1&2&3\end{smallmatrix}&
      \parbox{7cm}{\normalfont\footnotesize
        \begin{align*}
          u^+_1 & = u^-_4, & u^+_2 &= u^-_6,  & u^+_3 & = u^-_5,  \\
          u^+_4 & = u^-_1, & u^+_5 &= u^-_2, & u^+_6 & = u^-_3.
        \end{align*}
      }
    \end{array}
  \end{equation*}
  \caption{Permutation and parameter assignment for the BCFW decomposition
    of the six-point $\mathrm{NMHV}$ amplitude}
  \label{fig:bcfw}
\end{figure}
Compatibility between the three channels is achieved by imposing all conditions
simultaneously. Doing so, there are three remaining degrees of freedom:
choosing $(u^+_3,\,u_5^+,\,u_6^+) := (a,b,c)$ as parameters, one finds
\begin{align}
  \label{eq:bcfw_def_sol}
   (u_1^+,\,u_1^-)&= (c, b ),&
   (u_2^+,\,u_2^-)&= (a, b ),&
   (u_3^+,\,u_3^-)&= (a, c ),\nln
   (u_4^+,\,u_4^-)&= (b, c ),&
   (u_5^+,\,u_5^-)&= (b, a ),&
   (u_6^+,\,u_6^-)&= (c, a ).
\end{align}
We stress that the deformation described above is obtained by imposing Yangian
invariance on each on-shell graph individually. We will see in the
next paragraph how this result compares to the other approach.

%%%%%%%%%%%%%%%%%%%%%%%%%%%%%%%%%%%%%%%%
\paragraph{Residues of the top-cell} 

The boundaries of an undeformed on-shell
graph can be obtained by removing a single edge from the graph. However, only
specific internal edges can be removed consistently~\cite{ArkaniHamed:2012nw}.
For the top-cell graph of $\mathcal{A}_{6,3}$, the removable edges are
highlighted in \figref{fig:6pttopcell_bdaries}.

Taking a codimension-one boundary of a graph corresponds to calculating a
residue of the Grassmannian integral. The lack of meromorphicity of the deformed
integrand makes the task of taking the residue nontrivial, since one has to take
care of the branch-cut structure. It is, however, always possible to force the
integral associated with the removal of a removable edge to be meromorphic: the
additional condition one has to impose is the vanishing of the central charge
associated with that particular edge. A rigorous derivation of the sufficiency
of this condition can be found in \appref{app:grassm}.

Yangian invariance of the top-cell corresponds to the  permutation and
linear system reported in \figref{fig:tc_uc}.
\begin{figure}
  \centering
  \begin{equation*}
    \begin{array}[ht]{c|c|c}
      \text{graph} & \text{permutation} & u\pm \zvar\\
      \hline &&\\
      \includegraphicsboxex[scale=.6]{Fig6ptnmhv_topcell.mps} &
      % (456123)
      \begin{smallmatrix}1&2&3&4&5&6\\\downarrow&\downarrow&\downarrow&\downarrow%
        &\downarrow&\downarrow\\4&5&6&1&2&3\end{smallmatrix}&
      \parbox{7cm}{\normalfont\footnotesize
        \begin{align*}
          u^+_1 &= u^-_4, & u^+_2 & = u^-_5, & u^+_3 & = u^-_6, \\
          u^+_4 &= u^-_1, & u^+_5 & = u^-_2, & u^+_6 & = u^-_3.
        \end{align*}
      }
    \end{array}
  \end{equation*}
  \caption{Permutation and parameter assignment for the top-cell
    on-shell graph for the six-point $\mathrm{NMHV}$ amplitude.}
  \label{fig:tc_uc}
\end{figure}
Enforcing the vanishing of central charges on all three edges to be removed
amounts to setting 
\[
  \label{eq:topc_uc}
  u_5^+\,=u_4^+ \ ,\qquad u_2^+\,=u_3^+ \ , \qquad u_1^+\,=u_6^+ \ .
\]
Consistently, the combination of the conditions reported in \tabref{fig:tc_uc}
and in \eqnref{eq:topc_uc} is equivalent to \eqnref{eq:bcfw_def_sol}. 

It is possible
to check that also the alternative BCFW decomposition -- 
involving graphs 1, 3 and 5 of \figref{fig:6pttopcell_bdaries}, 
which corresponds to the removal of
the green lines -- admits a nontrivial deformation compatible with the
codimension-one boundaries of the top-cell. Imposing vanishing of the central
charges flowing on the red and green lines \figref{fig:6pttopcell_bdaries}
simultaneously leads to a linear system forcing all evaluation parameters to be
equal and all central charges to be zero: Yangian invariance can only be
realised in the undeformed case, as there are not enough degrees of freedom
available. 

The fact that the different on-shell graphs constituting a particular 
BCFW decomposition of a given tree-level NMHV amplitude 
admit a nontrivial common deformation is peculiar to the six-point case. 
Already for $7$-point NMHV, imposing the
compatibility for the deformations of the six on-shell graphs representing the
six BCFW channels gives only the trivial solution: all central charges are
forced to vanish and all evaluation parameters are equal. This is to be
expected, since for tree amplitudes the number of BCFW channels for fixed $k$
grows roughly quadratically with $n$, whereas the number of free deformation
parameters is of order $n$.  

Therefore -- with the exception of MHV amplitudes and a few low-multiplicity
examples in the non-MHV sector -- it is not possible to express a deformed
amplitude as a sum of deformed objects which are Yangian-invariant individually
because the constraints arising from demanding compatibility between the
Yangian-invariant constituent graphs lead to the trivial deformation: all
central charges are fixed to zero and all spectral parameters are equal.

%%%%%%%%%%%%%%%%%%%%%%%%%%%%%%%%%%%%%%%%%%%%%%%%%%%%%%%%%%%%%%%%%%%%%%%%%%%%%%%%
\subsection{The central charge as a regulator for the four-point one-loop amplitude}
\label{sec:oneloop}

In refs.~\cite{Ferro:2012xw,Ferro:2013dga} the possibility of using the
deformation parameters as regulators for loop amplitudes was investigated.  The
authors considered and computed a deformed on-shell graph which -- in the undeformed limit
-- corresponds to the four-point one-loop MHV amplitude. 
The result was manifestly finite, but it turned out not to preserve Yangian symmetry
essentially because the latter was not taken into account in the construction.

Our focus is to investigate the possibility of using a
Yangian-preserving deformation as a means to regulate loop amplitudes. The case
of study will again be the on-shell graph that corresponds to the four-point
one-loop MHV amplitude in the undeformed limit%
\footnote{%
  The impossibility of conciliating consistently BCFW decomposition and
  deformation of on-shell diagrams -- investigated in \subsecref{sec:6nmhv} --
  does not pose any problem here, since in the undeformed limit this graph is
  the only graph corresponding to the four-point one-loop amplitude. }. %
We evaluate the integral corresponding to a Yangian-invariant deformation and
study its finiteness. As we shall see, the result we find 
is rather unexpected and curious.  

\begin{figure}
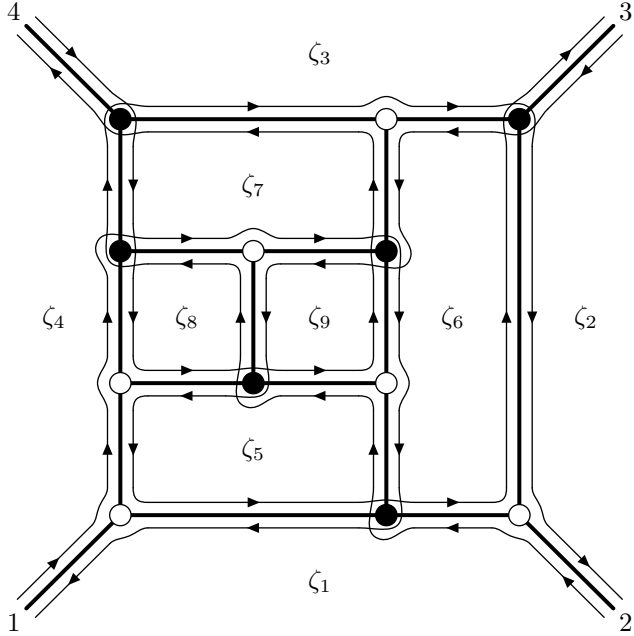

  \centering
  \includegraphicsboxex{4pts1loop.mps}
  \caption{Four-point one-loop on-shell graph}
  \label{fig:4pt1loop}
\end{figure}
The on-shell graph associated with the four-point one-loop amplitude is drawn
in \figref{fig:4pt1loop}, where we labelled the face shifts defined above
\eqnref{eq:faceshift2} by $\zeta_1\ldots\zeta_9$. Imposing Yangian
invariance forces all external charges to vanish, which corresponds to the
equality of all external face shifts $\zeta_1\ldots\zeta_4$. The face shifts
are related to the evaluation parameters $u_i$ of the external legs via  
 \begin{align}
  \label{eq:fv_ours}
    \zeta_1 = \zeta_2 = \zeta_3 = \zeta_4 &= \Delta, & \zeta_5 &=\Delta + u_3 - u_1,\nln
    \zeta_6 &= \Delta + u_3 - u_2, &  \zeta_7 &= \Delta + u_4 - u_2, \nln
    \zeta_8 &= \Delta + u_4-u_1, & \zeta_9 &= \Delta+u_4-u_1+u_3-u_2,
\end{align}
as can be easily obtained by following the parameters $u^\pm$ along the
permutation lines in \figref{fig:4pt1loop}. As already stated, this assignment
of face shifts here preserves Yangian invariance and differs from the one
studied in refs.~\cite{Ferro:2012xw,Ferro:2013dga}.

\begin{figure}
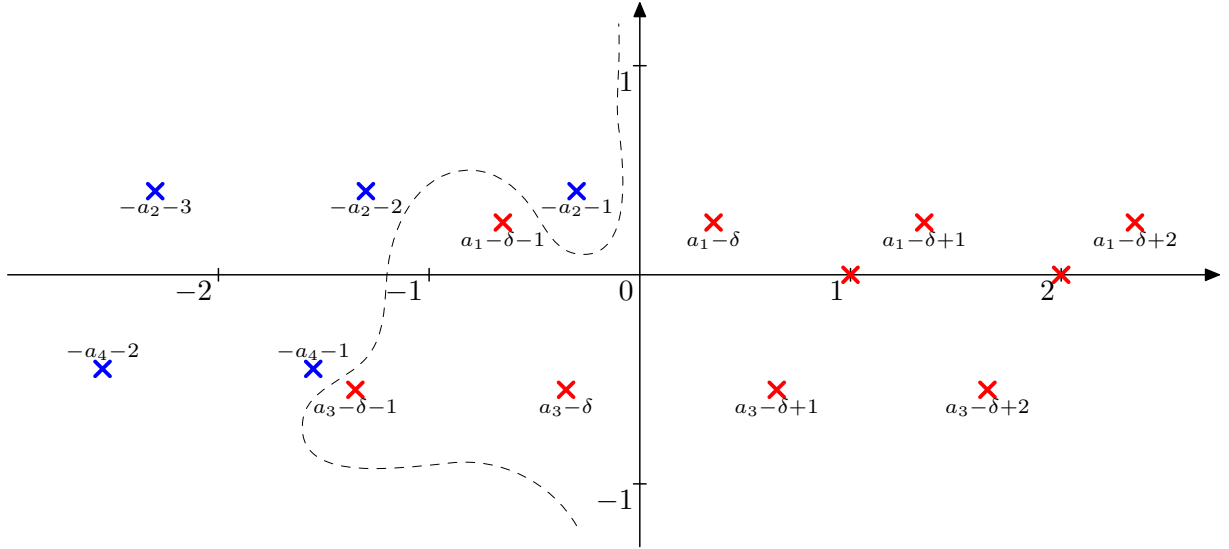

  \centering
  \includegraphicsboxex[width=\textwidth]{FigMBpoles.mps}
  \caption{Poles in the MB representation of the amplitude. The dashed line is
    the integration contour.}
    \label{fig:MBContour}
\end{figure}

The technique to construct the loop integral in terms of local variables was
described by the authors of refs.~\cite{Ferro:2012xw,Ferro:2013dga}.  As computed
in these papers, the deformed four-point one-loop amplitude can be expressed in
terms of a massless box integral with a deformed integrand, whose deformation is
reminiscent of analytic regularisation:
\[
  \label{eq:integrand}
  \mathcal{A}_{4,2}^{\text{1-loop}} (\{a_i\};s,t) =
  s\,t\,\mathcal{A}_{4,2}^{\text{tree}}\,\cdot\,I_{\text{box}}(\{a_i\};s,t),
\]
where $s=2 k_1\cdot k_2$ and $t = 2 k_2\cdot k_3$ are Mandelstam variables and
the deformed one-loop box integral reads%
\footnote{%
  A curious fact is that no spinor quantities arise in the integral after
  Yangian invariance is imposed, when for general face variables they do. This
  suggests that the integral may be performed in dimensional regularisation, if
  required.  }%
\[
  \label{eq:box_fv_ours}
  I_{\mathrm{box}}(\{a_i\};s,t) = \int\frac{\mathrm{d}^4q}{(q^2)^{1+a_1} [(q+k_1)^2]^{1+a_2}
    [(q+k_1+k_2)^2]^{1+a_3} [(q-k_4)^2]^{1+a_4}}.
\]
Here the variables $a_i$ are the following combinations of the evaluation parameters $u_k$
\[a_1 = u_4 - u_1,\qquad
a_2 = u_3 - u_4,\qquad
a_3 = u_2 - u_3,\qquad
a_4 = u_1 - u_2,\]
which implies $\delta :=\sum_i a_i = 0$.  However, for the following
calculation it is advisable to keep the dependence on $\delta$ explicit, as it
will provide easier access to the singular behaviour later on. Introducing
Feynman parameters $x_k$, the integral can be rewritten as
\begin{align}
  I_{\mathrm{box}}& = \frac{\gammafn(4+\sum_i
    a_i)}{\prod_{j=1}^4 \gammafn(1+a_j)} \int\mathrm{d}^4q\int_0^1
  \prod_{k=1}^4\Bigl[\mathrm{d}x_k\,x_k^{a_k}\Bigr]\,
  \delta\Bigl(1-\sum_{i=1}^4 x_i\Bigr)\times \notag\\
  &\qquad\qquad\times\Bigl[\Bigl(\sum_{i=1}^4x_i\Bigr)q^2
  +2q\cdot(x_2k_1+x_3k_1+x_3k_2-x_4k_4)
  +x_3 \,s\Bigr]^{-4-\sum_{i=1}^4 a_i} \notag\\
  &=\,(-1)^\delta\frac{\gammafn(2+\delta)}{\prod_{j=1}^4
    \gammafn(1+a_j)} \int_0^1
  \prod_{k=1}^4\Bigl[\mathrm{d}x_k\,x_k^{a_k}\Bigr]\,
  \delta\Bigl(1-\sum_{i=1}^4 x_i\Bigr) \, \Bigl[x_1 x_3\,s +
  x_2 x_4 \,t\Bigr]^{-2-\delta}.
  \label{eq:box_feymn_pars}
\end{align}
In order to evaluate the integral, we employ the Mellin--Barnes
representation,%
\footnote{We are grateful to Jan Plefka and Radu Roiban for
  discussions of this point.} 
which yields
\begin{align}
  \label{eq:MB_init}
  I_{\mathrm{box}}
  & =
  \frac{1}{2\pi i}\frac{(-1)^{\delta}}{\prod_{j=1}^4 \gammafn(1+a_j)}\times
  \notag\\
  & \qquad\times\int_{-i\infty}^{i\infty}\!\!\mathrm{d}z
  \int_0^1\prod_{k=1}^4\Bigl[\mathrm{d}x_k\,x_k^{a_k}\Bigr]\,
  \delta\Bigl(1-\sum_i x_i\Bigr)
  \frac{t^{z}}{s^{2+\delta+z}}
  \frac{(x_2x_4)^{z}}{(x_1x_3)^{z+2+\delta}}
  \gammafn(-z) \gammafn(z+2+\delta) \notag\\[4pt]
  & = 
  \frac{1}{2\pi i}\frac{(-1)^\delta}{\gammafn(-\delta)\prod_{j=1}^4 \gammafn(1+a_j)}
  \frac{1}{s^{2+\delta}}\times\notag\\
  &\qquad\times 
  \int_{-i\infty}^{i\infty}\!\!\mathrm{d}z\,\biggl(\frac{t}{s}\biggr)^z
  \gammafn(1+a_2+z)
  \,\gammafn(1+\delta-a_1-a_2-a_3+z)\times\notag\\
  &\qquad\qquad\qquad\times
  \gammafn(a_1-1-\delta-z)\,
  \gammafn(a_3-1-\delta-z)\,\gammafn(2+\delta+z)\,\gammafn(-z),
\end{align}
where the contour of integration is drawn in \figref{fig:MBContour}.  It
separates the poles originating in the gamma functions with argument
$\gammafn(\dots -z)$ (right poles) from the poles of $\gammafn(\dots + z)$ (left
poles). 
In \eqnref{eq:MB_init} one can see why it was useful to keep the parameter $\delta$
explicit. Instead of being a regulator, it serves as an auxiliary variable in
order to investigate the convergence of the integral. Yangian invariance
requires $\delta=0$; therefore one would na\"{i}vely conclude that the result is
zero because of the prefactor $1/\gammafn(0)$.  However, it is still possible for
the integral to be nonvanishing if the integral diverges as
$1/\delta$. We will investigate this issue.

\begin{figure}
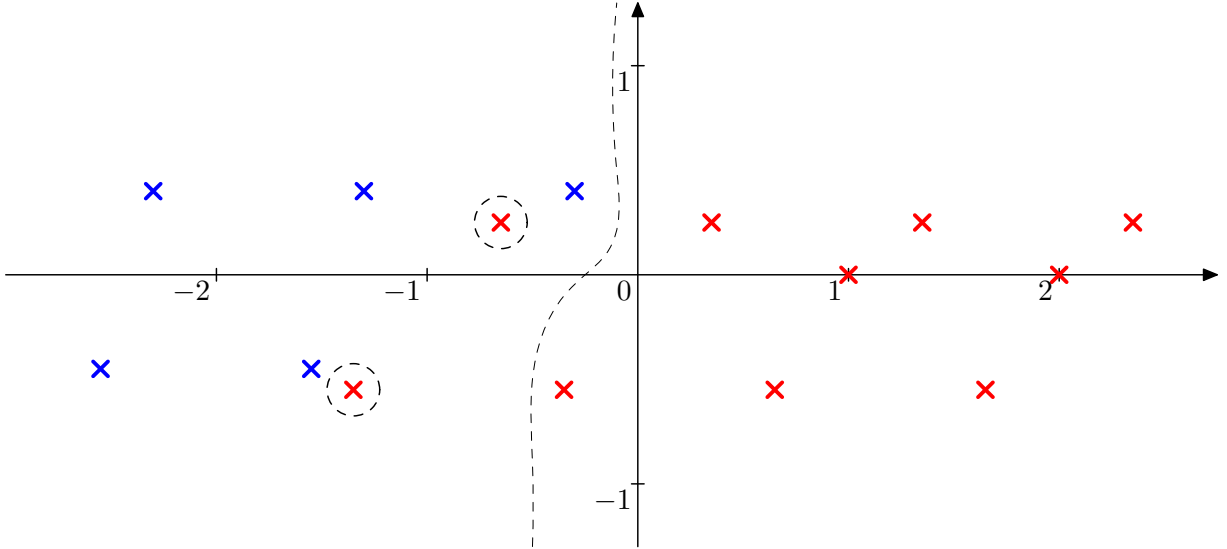

  \centering
  \includegraphicsboxex[width=\textwidth]{FigMBcontour.mps}
  \caption{Contour of integration with possible pinchings disentangled; the two
    dashed circles represent residue integrals around the poles at
    $a_3-\delta-1$ and $a_1-\delta-1$, respectively.}
  \label{fig:contour}
\end{figure}

The Mellin--Barnes integral diverges whenever a left and a right pole become
coincident -- a situation called \emph{pinching} of poles. In order to extract
the possibly divergent contribution, one has to split the contour of
integration such that the right poles giving rise to the pinching are moved to
the left. Accordingly, the contour integral splits into a convergent part
arising from a new contour parallel to the imaginary axis and the residue
integrals around the problematic right poles as drawn in
\figref{fig:contour}. 
Computing the two residues at $a_1-1-\delta$ and $a_3-1-\delta$ leads to
\[
  \label{eq:MB_2res}
  \begin{aligned} 
    I_{\mathrm{box}} = \frac{1}{s^{2+\delta}\gammafn(-\delta)} \Biggl\{
    &\biggl(\frac{t}{s}\biggr)^{a_1-1-\delta} \frac{\gammafn(1+\delta-a_1)
      \gammafn(-\delta+a_1+a_2)\gammafn(-a_1+a_3)
      \gammafn(-a_2-a_3)}{\gammafn(1+a_2)\gammafn(1+a_3)\gammafn(1+\delta-a_1-a_2-a_3)} + \\
    + &\biggl(\frac{t}{s}\biggr)^{a_3-1-\delta} \frac{\gammafn(1+\delta-a_3)
      \gammafn(-\delta+a_2+a_3)\gammafn(a_1-a_3)
      \gammafn(-a_1-a_2)}{\gammafn(1+a_1)\gammafn(1+a_2)\gammafn(1+\delta-a_1-a_2-a_3)}
    \Biggr\}.
  \end{aligned}
\]
Imposing $\delta=0$ in \eqnref{eq:MB_2res} clearly makes the integral vanish
for generic values of
$a_1,a_2,a_3$: 
\[
  \label{eq:integral_zero}
  I_{\mathrm{box}} = 0.
\]
If this was the complete result, there would be a contradiction to the usual
calculation of an undeformed four-point box integral.  In particular, the
undeformed box integral suffers from infrared divergences, whereas the above
equality would imply it to vanish.

However, a more careful analysis shows that for specific values of $a_1,a_2,a_3$
the result may be nonzero. For example, when $a_1+a_2 = 0$ and $a_2+a_3=0$, the
$1/\gammafn(-\delta)$ is cancelled in both terms of \eqnref{eq:MB_2res}. Hence,
a double pole appears originating from two $\gammafn$'s in the numerator.  This
suggests that the result is not identically zero, but rather a distribution with
simultaneous support at $a_1+a_2=0$ and $a_2+a_3=0$.
The derivation is subtle and requires much care. 
Using a heuristic derivation we find
\[
  \label{eq:integral_delta}
  I_{\mathrm{box}}\Bigr\vert_{\delta=0} = - \delta(a_1+a_2)\, \delta(a_2+a_3)\, \frac{1}{s \, t} \biggl(
  \frac{t}{s} \biggr)^{a_1} \frac{\sin (a_1 \pi) }{ a_1 } \ .
\]
It is straightforward to check that this result is invariant under the exchange
$s \,\leftrightarrow\, t$, provided that one exchanges also $a_1
\,\leftrightarrow\, a_4, a_2\, \leftrightarrow\, a_3$, as can be deduced
directly from the integral of \eqnref{eq:box_fv_ours}. 
Moreover, the result in the undeformed case, $a_i=0$,
is proportional to $\delta^2(0)$ which could be argued to 
agree qualitatively with the $\log^2$ divergence of the undeformed integral.
We refer to \appref{app:integral} for more details on our derivation.

%%%%%%%%%%%%%%%%%%%%%%%%%%%%%%%%%%%%%%%%%%%%%%%%%%%%%%%%%%%%%%%%%%%%%%%%%%%%%%%%
\subsection{Discussion}

Let us briefly summarise our findings: 
In order to define deformed tree-level scattering amplitudes, there is only one
object which can be deformed in a Yangian-invariant way without
any further constraints: the top-cell. Integrating over the integrand
corresponding to a deformed (tree-level) top-cell will lead to deformed MHV
amplitudes. 

Non-MHV tree-level amplitudes are built as a sum of Yangian invariant
constituent graphs. While deforming each individual constituent graph works as
for the top-cell, imposing compatibility between all graphs constrains the
deformation parameters. While a few low-multiplicity examples still allow a
deformation, the majority of non-MHV amplitudes cannot be deformed in this
framework. 

A possible way out would be the construction of Yangian-invariant scattering
tree amplitudes in the non-MHV sector from constituent graphs which violate
Yangian invariance individually, but sum up to an Yangian-invariant
combination. It is not clear, whether this approach can be successful. Clearly, the
formalism of deformed on-shell graphs explored in this paper is not applicable. 

For what concerns amplitudes beyond tree-level, we restricted our study to the
four-point one-loop case. In the undeformed case, this amplitude can be
expressed via a single on-shell graph taking the r\^ole of the top-cell. The
computation of the integral corresponding to the Yangian-invariant deformation
of this on-shell graph yields a curious result: na\"{i}vely it would appear to be
zero -- which is trivially Yangian invariant. A more careful analysis 
suggests it to be a distribution with singular support.

For amplitudes of higher loop order and higher multiplicity, there is no known
notion of an on-shell diagram comparable to the top-cell. Thus the situation is
less clear than in the tree-level case.

%%%%%%%%%%%%%%%%%%%%%%%%%%%%%%%%%%%%%%%%%%%%%%%%%%%%%%%%%%%%%%%%%%%%%%%%%%%%%%%%
%%%%%%%%%%%%%%%%%%%%%%%%%%%%%%%%%%%%%%%%%%%%%%%%%%%%%%%%%%%%%%%%%%%%%%%%%%%%%%%%
\section{Conclusions}
\label{sec:conclusions}

In this paper, we describe a mechanism for constructing invariants 
of the Yangian algebra $\yang[\alg{su}(2,2|4)]$
by combining trivalent invariant building blocks and
maintaining invariance during the gluing procedure. 
The invariants are deformations
of the on-shell diagrams put forward in ref.~\cite{ArkaniHamed:2012nw}
which are invariants of $\yang[\alg{psu}(2,2|4)]$. 
Deformation refers to a continuous shift of the central charges 
of the external legs introduced in \cite{Ferro:2012xw,Ferro:2013dga}. 

The construction of deformed tree-level scattering amplitudes from deformed
on-shell graphs works straightforwardly only for MHV amplitudes. With a few
exceptions, the non-MHV amplitudes can not be deformed in the framework of
(manifestly) Yangian-invariant on-shell graphs.  

In ref.~\cite{Bargheer:2009qu} it was argued that once one considers the
holomorphic anomaly only, the complete S-matrix of $\superN=4$ sYM is an
invariant object as opposed to the constituent scattering amplitudes.  For the
deformed case we have argued that the interpretation of on-shell amplitudes as
constituents of an S-matrix is all but evident.  Nevertheless one could hope
that the assumption of an exactly invariant composite object may lead the way
to the identification of a suitable deformed S-matrix and its constituent
on-shell graphs.

Finally, in ref.~\cite{Ferro:2012xw,Ferro:2013dga} it was suggested to employ
the shifted helicities of the external legs as a regulator maintaining
superconformal invariance.  However, for the example of the one-loop amplitude,
Yangian invariance was spoilt.  If we stick with a Yangian-invariant
deformation of the integrand, the resulting expression for the four-point
one-loop amplitude will be zero for generic deformation parameters.  For special
deformations, however, the divergent behaviour of the integral is difficult to
access. It depends on the particular way the deformation parameters are
taken to zero.

%%%%%%%%%%%%%%%%%%%%%%%%%%%%%%%%%%%%%%%%
\paragraph{Acknowledgements}

The authors would like to thank
Claude Duhr, 
Livia Ferro,
To\-masz \L{}ukowski,
Carlo Meneghelli,
Jan Plefka,
Matthias Staudacher,
Cristian Vergu 
for discussions.

The work of NB and MR is partially supported
by grant no.\ 200021-137616 from the Swiss National Science Foundation.

%%%%%%%%%%%%%%%%%%%%%%%%%%%%%%%%%%%%%%%%%%%%%%%%%%%%%%%%%%%%%%%%%%%%%%%%%%%%%%%%
% Appendices
%%%%%%%%%%%%%%%%%%%%%%%%%%%%%%%%%%%%%%%%%%%%%%%%%%%%%%%%%%%%%%%%%%%%%%%%%%%%%%%%
\appendix

%-------------------------------------------------------------------------------

%%%%%%%%%%%%%%%%%%%%%%%%%%%%%%%%%%%%%%%%%%%%%%%%%%%%%%%%%%%%%%%%%%%%%%%%%%%%%%%%
%%%%%%%%%%%%%%%%%%%%%%%%%%%%%%%%%%%%%%%%%%%%%%%%%%%%%%%%%%%%%%%%%%%%%%%%%%%%%%%%
\section{Yangian invariance of deformed glued objects}
\label{app:yangian_inv}

In this appendix we provide the explicit calculation showing that an object
obtained by gluing two deformed Yangian invariants is Yangian invariant if
evaluation parameters and central charges are identified on the internal legs as
in \eqnref{eq:gluedobject}. Let us consider two objects,
\[
\mathcal{A}(1,\dots,m,I),\;\mathcal{B}(I,m+1,\dots,n) \ ,
\]
which are Yangian invariant, that is
\begin{align}
  \label{eq:yinvAB}
  \alg{J}^a\cdot\,\mathcal{A}\,&\equiv\,\Bigl[\sum_{i=1}^I \alg{J}^a_i \Bigr]
  \cdot \mathcal{A} = 0,\\
  \widehat{\alg{J}}^a\cdot\,\mathcal{A}\,&\equiv \Bigl[f^a_{\;b c}
  \sum_{i=1}^m\sum_{j=i+1}^I \alg{J}^b_i \,\alg{J}^c_j +
  \sum_{k=1}^I \,u_k\,\alg{J}^a_k \Bigr]\cdot \mathcal{A} = 0 \ ,
\end{align}
and analogously for $\mathcal{B}$.
Defining 
\[
  \mathrm{d}^{4|4}\xi := \frac{\mathrm{d}^2\lambda_I\,
  \mathrm{d}^2\tilde{\lambda}_I}{\mathrm{Vol}[\grp{GL}(1)]}\,\mathrm{d}^{0|4}\eta_I \ ,
\]
the glued object reads
\[
\mathcal{Y}(1,\dots,n) = 
\int \mathrm{d}^{4|4}\xi
\, \mathcal{A}
(1,\dots,m,I)\,\mathcal{B}(I,m+1,\dots,n)\,.
\]
It is invariant under the action of the level-zero generators for the same reason
as in the undeformed case (see the discussion after \eqnref{eq:intmeasure}),
provided that it is annihilated by the total central charge operator
$\alg{C}=\sum_{i=1}^n\alg{C}_i$. Combined with the Yangian invariance of
$\mathcal{A}$ and $\mathcal{B}$, this implies that the central charges attached
to the leg $I$ of $\mathcal{A}$ must be the same (up to a sign) as the central
charge attached to the leg $I$ of $\mathcal{B}$.

In order to derive the condition for gluing the evaluation parameters, it is
useful to note that the invariance of $\mathcal{Y}$ under the action of
level-zero
generators implies
\[
\label{eq:y_ibp}
\int
\mathrm{d}^{4|4}\xi\, \Bigl[\bigl(\alg{J}_I^a \mathcal{A}\bigr)
\,\mathcal{B} + \mathcal{A}\,\bigl(\alg{J}_I^a \mathcal{B}\bigr) \Bigr] \,=\,0.
\]
However, in order to fix the gluing condition for $u_I$, we have to consider
the action of a level-one generator on the glued object:
\begin{align}
  \widehat{\alg{J}}^a\cdot& \int
  \mathrm{d}^{4|4}\xi \,\mathcal{A}(1,\dots,m,I)\mathcal{B}(I,m+1,\dots,n)\nn\\
  =& f^a_{\;b c}
  \int\mathrm{d}^{4|4}\xi  \biggl[
  \sum_{i=1}^{m}\sum_{j=i+1}^m \Bigl( \alg{J}^b_i \alg{J}^c_j \mathcal{A} \Bigr)\mathcal{B} +
  \sum_{i=1}^{m}\sum_{j=m+1}^n \Bigl( \alg{J}^b_i \mathcal{A} \Bigr)
  \Bigl(  \alg{J}^c_j \mathcal{B} \Bigr) +\nln
  &\qquad\qquad\qquad\qquad+\sum_{i=m+1}^{n-1}\sum_{j=i+1}^n  \mathcal{A}
  \Bigl( \alg{J}^b_i \alg{J}^c_j \mathcal{B} \Bigr) 
  \biggr]
  +\sum_{k=1}^n u_k \alg{J}^a_k \int \mathrm{d}^{4|4}\xi\mathcal{A}\mathcal{B}\nln
  =& f^a_{\;b c}
  \int\mathrm{d}^{4|4}\xi  \biggl[
  \sum_{i=1}^{m}\sum_{j=i+1}^I \Bigl( \alg{J}^b_i \alg{J}^c_j \mathcal{A} \Bigr)\mathcal{B} -
  \sum_{i=1}^m \Bigl( \alg{J}^b_i \alg{J}^c_I \mathcal{A} \Bigr)\mathcal{B} +
  \sum_{i=1}^{m}\sum_{j=m+1}^n \Bigl( \alg{J}^b_i \mathcal{A} \Bigr)
  \Bigl(  \alg{J}^c_j \mathcal{B} \Bigr) +\nln
  &\qquad\qquad\qquad\qquad-\sum_{j=m+1}^n \mathcal{A} \Bigl( \alg{J}^b_I \alg{J}^c_j \mathcal{B} \Bigr) +
  \sum_{i=m+1}^{n-1}\sum_{j=i+1}^n  \mathcal{A} \Bigl( \alg{J}^b_i \alg{J}^c_j \mathcal{B} \Bigr)
  \biggr] + \sum_{k=1}^n  u_k \alg{J}^a_k \int \mathrm{d}^{4|4}\xi\mathcal{A}\mathcal{B}
\end{align}
In the second and fourth term in the last line one can move the $\alg{J}_I$ from
$\mathcal{A}$ to $\mathcal{B}$ (paying a minus sign) and vice versa
using~\eqref{eq:y_ibp}. Then the second, third and fourth term can be combined
into $(\alg{J}^b \cdot\mathcal{A})(\alg{J}^c\cdot \mathcal{B})$ which obviously
vanishes. The remaining first and fifth term together with the term containing
the evaluation parameters vanishes for all external legs leaving a
contribution from the internal line $I$ which reads
\[
  \label{eq:result}
  \widehat{\alg{J}}^a\cdot \int
  \frac{\mathrm{d}^2\lambda\,\mathrm{d}^2\tilde{\lambda}}{\mathrm{Vol}[\grp{GL}(1)]}
  \,\mathrm{d}^{0|4}\eta_I \,\mathcal{A}\,\mathcal{B} = \int
  \frac{\mathrm{d}^2\lambda\,\mathrm{d}^2\tilde{\lambda}}{\mathrm{Vol}[\grp{GL}(1)]}
  \,\mathrm{d}^{0|4}\eta_I
  \biggl[ u_I^\mathcal{A} \Bigl(\alg{J}^a_I \mathcal{A}\Bigr)\mathcal{B} +
  u_I^\mathcal{B}\mathcal{A} \Bigl(\alg{J}^a_I\mathcal{B}\Bigr)  \biggr]\,.
\]
Using the identity~\eqref{eq:y_ibp}, one can show that the above expression
vanishes if $u^\mathcal{A}_I = u^\mathcal{B}_I$. Therefore, superconformal and
dual superconformal invariance imply that the evaluation parameters on internal
legs match.

In a similar way, it is also possible to show that, given a Yangian invariant,
the object obtained by identifying two adjacent legs and integrating over the
on-shell superspace of that leg is again a Yangian invariant. Specifically,
given an $(n+2)$-point Yangian invariant $\mathcal{Y}(1,\dots,n+2)$, the object
obtained via
\begin{equation}
  \label{eq:y_loop}
  \mathcal{Y}'(1,\dots,n) = \int \frac{\mathrm{d}^2\lambda_I\,
    \mathrm{d}^2\tilde{\lambda}_I}{\mathrm{Vol}[\grp{GL}(1)]}\,\mathrm{d}^{0|4}\eta_I
  \,\mathcal{Y}(1,\dots,n,I,J)\vert_{J=I}
\end{equation}
is again Yangian invariant. The only nontrivial check is invariance under
level-one generators; a sketch of the invariance (in the undeformed case) is as
follows
\begin{align}
  \widehat{\alg{J}}^a\cdot& \int \mathrm{d}^{4|4}\xi \,\mathcal{Y}
  (1,\dots,n,I,J)\vert_{I=J}\nn\\
  =& f^a_{\;b c} \int\mathrm{d}^{4|4}\xi\Bigl[ \sum_{i=1}^{n-1}\sum_{j=i+1}^n
  \Bigl( \alg{J}^b_i \alg{J}^c_j \,\mathcal{Y}
  (1,\dots,n,I,J)\Bigr)\Bigr]\Bigr\vert_{I=J}\nn \\
  =& - f^a_{\;b c} \int\mathrm{d}^{4|4}\xi\Bigl[ \alg{J}_I^b \alg{J}_J^c
  \mathcal{Y} + \sum_{i=1}^{n} \Bigl( \alg{J}^b_I \alg{J}^c_I 
  \mathcal{Y}\Bigr) + \sum_{i=1}^{n} \Bigl( \alg{J}^b_I \alg{J}^c_J 
  \mathcal{Y}
  \Bigr)\Bigr]\Bigr\vert_{I=J} = % f^a_{\;b c}
  % \int\mathrm{d}^{4|4}\xi\Bigl[ \alg{J}_I^b \alg{J}_J^c
  % \mathcal{Y} \Bigr]\Bigr\vert_{I=J}
  \nn \\
  =& f^a_{\;b c} \int\mathrm{d}^{4|4}\xi\Bigl[ \alg{J}_I^b \alg{J}_I^c
  \mathcal{Y} \Bigr]\Bigr\vert_{I=J} \propto f^a_{\;b c} f^{b c}_{\;d}
  \int\mathrm{d}^{4|4}\xi\,\Bigl[ \alg{J}_J^d \mathcal{Y} \Bigr]\Bigr\vert_{I=J}
\end{align}
which vanishes for $\alg{psu}(2,2|4)$. In the derivation we implicitly used the
invariance under the level-zero generators and the ``integration-by-parts''
identity
\[
\label{eq:y_ibp_2}
\int
\mathrm{d}^{4|4}\xi\, \Bigl[\alg{J}_I^a \mathcal{Y} (\dots,I,J) +
\alg{J}_J^a \mathcal{Y} (\dots,I,J)\Bigr]_{I=J} \,=\,0
\]
similar to \eqnref{eq:y_ibp}. The generalisation to the deformed case is
straightforward and leads to the same conditions for $\zvar$'s and $u$'s as
before.

%%%%%%%%%%%%%%%%%%%%%%%%%%%%%%%%%%%%%%%%%%%%%%%%%%%%%%%%%%%%%%%%%%%%%%%%%%%%%%%%
%%%%%%%%%%%%%%%%%%%%%%%%%%%%%%%%%%%%%%%%%%%%%%%%%%%%%%%%%%%%%%%%%%%%%%%%%%%%%%%%

\section{Grassmannian integrals from on-shell graphs}
\label{app:grassm}

One can associate a Grassmannian integral to any on-shell graph. The integral
is most conveniently expressed in terms of the face variables $f_i$ introduced
in \subsecref{sec:grassm}.  As already stated there, the Grassmannian
integral reads
\[
  \label{eq:onshell_app}
  \mathcal{I}_{\mathrm{graph}}\,=\, \int \biggl[\prod_{i=1}^{n_{\mathrm{F}}-1}\frac{\mathrm{d}f_i}{f_i}\biggr]
  \, \delta^{2k} \Bigl( \sum_{a=1}^n C_{r a}(f_i) \tilde{\lambda}_a \Bigr)\,
  \delta^{2(n-k)}\Bigl( \sum_{s=1}^{n-k} C_{s a}(f_i) \lambda_a \Bigr)\,
  \delta^{0|4k}  \Bigl( \sum_{a=1}^n C_{r a}(f_i) \tilde{\eta}_a \Bigr) \ ,
\]
where $C_{r a}(f_i)$ are elements of a $k\times n$ matrix that represents a point
in $G(k,n)$. The expression of $C_{r a}$ in terms of the face variables $f_i$ is
known (see ref.~\cite{ArkaniHamed:2012nw} for the precise construction).  

\begin{figure}
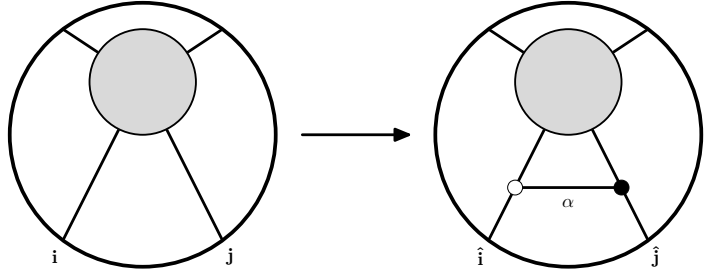

  \centering
  \includegraphicsboxex[scale=.7]{Fig_bridge.mps}
  \caption{BCFW bridge construction}
  \label{fig:bcfw_bridge}
\end{figure}

What is more interesting from our point of view is that the integral can be
expressed in terms of a different set of variables $\alpha$ which have a
precise physical interpretation: they are related to the BCFW shift associated
to adding a so-called BCFW bridge as depicted in \figref{fig:bcfw_bridge}.
The momentum flowing along the internal line is fixed to be proportional to
$\lambda_i \tilde{\lambda}_j$ by the deltas of the two additional vertices while the
external momenta are modified
\[
  \label{eq:bcfw_bridge}
  \begin{aligned}
    &\lambda_{\hat{i}} \,=\, \lambda_i,&&\qquad\lambda_{\hat{j}} \,=\,\lambda_j + \alpha\,\lambda_i,\\
    &\tilde{\lambda}_{\hat{i}} \,=\, \tilde{\lambda}_i - \alpha\,\tilde{\lambda}_j,
    &&\qquad\tilde{\lambda}_{\hat{j}} \,=\, \tilde{\lambda}_j.
  \end{aligned}
\]
In ref.~\cite{ArkaniHamed:2012nw}, it was demonstrated that the integral
associated with an on-shell graph can be expressed in terms of these edge
variables, the measure of integration being
$\prod\frac{\mathrm{d}\alpha}{\alpha}$. Moreover, it is also possible to show
that the removal of an edge $I$ corresponds to taking the residue around
$\alpha_I=0$. Not all the edges of a graph are removable, but the ones that are
can be identified from the permutation encoded by the graph. 

\begin{figure}
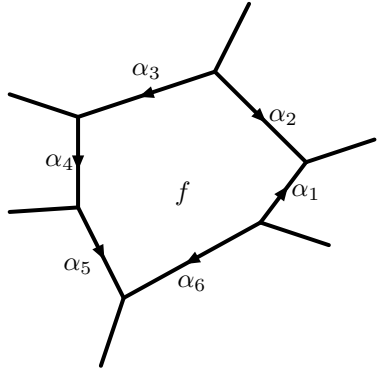

  \centering
  \includegraphicsboxex{Fig_facev_bound.mps}
  \caption{Face variable in terms of edge variables: here $f=\frac{\alpha_1
      \alpha_3 \alpha_4 \alpha_5}{\alpha_2 \alpha_6}$}
  \label{fig:facev_bound}
\end{figure}

It is possible to relate a face variable to the edge variables of the adjoining
edges. In order to do so, one must introduce a specific orientation of the
edges of the graph, called perfect orientation. Instead of describing the
technicalities of perfect orientation, let us stick with the result: a face
variable is given by the product of all the adjacent edge variables with
counterclockwise orientation divided by the product of all the adjacent edge
variables with clockwise orientation. An example is given in
\figref{fig:facev_bound}.

Considering the removal of an edge in the deformed integral, it is
evident that the measure in general will not be meromorphic in $\alpha$.
However, as pointed out in \eqnref{eq:faceshift}, the deformed measure in
terms of the face variables reads
\[
\prod \frac{\mathrm{d}f_i}{f_i}\;\to\; \prod \frac{\mathrm{d}f_i}{f_i^{1+\zeta_i}}.
\]
Focusing on one particular edge $I$, the related edge variable will appear in
the measure as $\mathrm{d}\alpha_I/\alpha_I^{1+\zeta_i-\zeta_j}$, where
$i$ and $j$ are the adjacent faces, since $\alpha_I$ appears in the numerator for
one face variable and in the denominator for the other. The measure will be of
the form $\frac{\mathrm{d}\alpha_I}{\alpha_I}$ if the difference of the face
shifts of the two adjacent faces is zero: this is equivalent to no central
charge flowing along the edge $I$.

%%%%%%%%%%%%%%%%%%%%%%%%%%%%%%%%%%%%%%%%%%%%%%%%%%%%%%%%%%%%%%%%%%%%%%%%%%%%%%%%
%%%%%%%%%%%%%%%%%%%%%%%%%%%%%%%%%%%%%%%%%%%%%%%%%%%%%%%%%%%%%%%%%%%%%%%%%%%%%%%%

\section{One-loop integral singular support}
\label{app:integral}

Here, we would like to argue
that the deformed one-loop box integral \eqnref{eq:box_fv_ours}
\[
  I_{\mathrm{box}}(\{a_i\};s,t) = \int\frac{\mathrm{d}^4q}{(q^2)^{1+a_1} [(q+k_1)^2]^{1+a_2}
    [(q+k_1+k_2)^2]^{1+a_3} [(q-k_4)^2]^{1+a_4}}
\]
in the form \eqnref{eq:MB_2res}
is expressed as a distribution with singular support, as in
\eqnref{eq:integral_delta}.

Let us draw an analogy with the case of a single Dirac delta. One can
define the Dirac delta function as the weak limit of the delta sequence
\begin{equation}
  \label{eq:deltaseq}
  \eta_\epsilon (x) := \frac{1}{2\pi i }
  \biggl[ \frac{1}{x - i \epsilon} - \frac{1}{x + i \epsilon} \biggr]
  = \frac{1}{\pi} \frac{\epsilon}{\epsilon^2 + x^2}\,,
\end{equation}
where $\eta_\epsilon$ satisfies the following limits:
\begin{alignat}{2}
\label{eq:eta_lim_1}
  \lim_{\epsilon \to 0}\,\eta_\epsilon (x) &= 0 \qquad &\text{for } x\neq 0,\\
\label{eq:eta_lim_2}
  \lim_{\epsilon \to 0}\,\epsilon \, \eta_\epsilon (c \epsilon) &=
  \frac{1}{\pi (c^2+1)} \qquad &\text{for } c\neq 0.
\end{alignat}
It is possible to show now that the function $I_{\mathrm{box}}$ defined in
\eqnref{eq:MB_2res} has a similar behaviour in the limit $\delta \to 0$. We already
saw that the limit $\lim_{\delta \to 0} I_{\mathrm{box}} = 0$ for generic
$a_1,a_2,a_3$; we want now to probe the region where $a_1\sim a_3\sim (-a_2)$.
We therefore consider $I_{\mathrm{box}}$ with $\sum_ia_i\propto\delta\neq 0$,
study the region where
\[
a_1+a_2 \sim \delta \ ,\qquad a_1 - a_3 \sim \delta \ ,
\]
and then take the limit $\delta \to 0$, in analogy with the limit taken in
\eqnref{eq:eta_lim_2}. 

Let us then rewrite the result of \eqnref{eq:MB_2res} as a function of
$a_1,a_2,a_3,a_4$, in order to obtain
\begin{equation}
  \label{eq:MB_2res_a}
  \begin{aligned} 
    I_{\mathrm{box}} = &\frac{1}{s^{2+a_1+a_2+a_3+a_4}\gammafn(-a_1-a_2-a_3-a_4)}\times\\
    &\times\Biggl\{ \biggl(\frac{t}{s}\biggr)^{-1 - a_2 - a_3 - a_4}
    \frac{\gammafn(1+a_2+a_3+a_4) \gammafn(-a_3-a_4)\gammafn(-a_1+a_3)
      \gammafn(-a_2-a_3)}{\gammafn(1+a_2)\gammafn(1+a_3)\gammafn(1+a_4)} + \\
    &\quad+ \biggl(\frac{t}{s}\biggr)^{-1-a_1-a_2-a_4} \frac{\gammafn(1+a_1+a_2+a_4)
      \gammafn(-a_1-a_4)\gammafn(a_1-a_3)
      \gammafn(-a_1-a_2)}{\gammafn(1+a_1)\gammafn(1+a_2)\gammafn(1+a_4)} \Biggr\} \ .
  \end{aligned}
\end{equation}
We subsequently set
\begin{equation}
  \label{eq:as_values}
  a_1 = a + d_1 \delta, \qquad    a_2 =-a + d_2 \delta, \qquad
  a_3 = a + d_3 \delta, \qquad    a_4 =-a + d_4 \delta,
\end{equation}
with $\sum_i a_i = d_{1234} \,\delta$,%
\footnote{ $d_{1234}$ is just an overall scale for the $d_i$'s; we can (and
  will) normalise them so that $d_{1234}=1$. } %
where $d_{ij\dots} := (d_i + d_j + \dots)$.  The limit we investigate is then
\[ 
\label{eq:app_limit}
\lim_{\delta\to 0} \;\delta^2\,I_{\mathrm{box}}\bigl(a+ d_1\delta,-a+ d_2\delta,a+
d_3\delta,-a+ d_4\delta\bigr).
\]
We stress the logic we are trying to follow here. We saw that $I_{\mathrm{box}}$
is vanishing for generic configurations of $a_i$'s, see \eqnref{eq:integral_zero}.
We are now treating the parameter $\delta$ as the $\epsilon$ of the delta
sequence, \eqnref{eq:eta_lim_1} and investigating the region where
$a_1+a_2\sim\delta$, $a_1-a_3\sim\delta$, in analogy with the study of the delta
sequence expressed in \eqnref{eq:eta_lim_2} in the region $x\sim\epsilon$.

The result of the limit in \eqnref{eq:app_limit} is
\begin{equation}
  \label{eq:lim_1}
  \begin{aligned}
    \lim_{\delta\to 0}\;\delta^2 I &= -\frac{1}{s \, t} \frac{(t/s)^{a}}{\gammafn(1+a)
      \gammafn(1-a)} \frac{(d_{1234})^2}{d_{12} \, d_{23} \, d_{34} \, d_{41}} \\
    & = -\frac{1}{s \, t}\frac{(t/s)^{a}}{\gammafn(1+a) \gammafn(1-a)}
    \biggl[
    \frac{1}{d_{12} \, d_{23}} + \frac{1}{d_{23} \, d_{34}} +
    \frac{1}{d_{34} \, d_{41}} + \frac{1}{d_{41} \, d_{12}}
    \biggr].
  \end{aligned}
\end{equation}
We can always rescale $\delta$, so we set $d_{1234}=1$. This leads to 
\begin{equation}
  \label{eq:lim_2}
  \lim_{\delta\to 0}\;\delta^2 I_{\mathrm{box}} = \frac{1}{s \, t}
  \frac{ (\frac{t}{s})^a }{\gammafn(1-a) \gammafn(1+a)}
  \frac{1}{ (d_{12} - 1) (d_{12}) (d_{23} - 1) (d_{23})},
\end{equation}
a result which is similar to \eqnref{eq:eta_lim_2}, but not equal; however,
the explicit form of the limit should depend on the direction along which the
limit of all $a_i$'s to be equal (up to a sign) is taken. It is possible to get
an expression with the same residues as \eqnref{eq:eta_lim_2} if we shift the
$a_i$'s differently, as
\begin{equation}
  \label{eq:lim_3}
  \lim_{\delta\to 0} \;\delta^2\,I_{\mathrm{box}}
  \biggl(
  a+ d_1\delta,\,
  -a+ \bigl( d_2 + \tfrac{i}{2} \bigr) \delta,\,
  a+ d_3\delta,\,
  -a+ \bigl( d_4 - \tfrac{i}{2} \bigr)\delta
  \biggr)
  \,:=\, \lim_{\delta\to 0} \;\delta^2\,I'_{\mathrm{box}},
\end{equation}
where we have defined a different shift for $a_2$ and $a_4$, keeping fixed the
condition $\sum_i a_i = d_{1234} \,\delta$.  The above equation leads to
\begin{equation}
  \label{eq:lim_4}
  \lim_{\delta\to 0} \;\delta^2\,I'_{\mathrm{box}} = - \frac{1}{s \, t}
  \biggl( \frac{t}{s} \biggr)^a
  \frac{16 \sin (a \pi) }{ a \pi } \frac{1}{ (4 d_{12}^2 + 1) (4 d_{23}^2 + 1) }
\end{equation}
which, considering also \eqnref{eq:eta_lim_2} (with $c \to c/2$), leads to
(cf.~\eqnref{eq:integral_delta})
\begin{equation}
  \label{eq:lim_5}
  I_{\mathrm{box}} = - \delta(a_1+a_2)\, \delta(a_2+a_3)\, \frac{1}{s \, t}
  \biggl( \frac{t}{s} \biggr)^{a_1}
  \frac{\sin (a_1 \pi) }{ a_1  }.
\end{equation}
There are however two remarks to be stressed. The first is that the choice of
shift in \eqnref{eq:lim_3} leads to a nonvanishing result with a clear
interpretation, but is \emph{ad hoc}.  The second is the argument leading to
\eqnref{eq:lim_5} is rather heuristic, based on the fact that the behaviour of
$I_{\mathrm{box}}$ in the limit $\delta\to0$ (and for a particular choice of
direction along with the $a_i$'s approach zero) is similar to the behaviour of a
delta sequence.

%%%%%%%%%%%%%%%%%%%%%%%%%%%%%%%%%%%%%%%%%%%%%%%%%%%%%%%%%%%%%%%%%%%%%%%%%%%%%%%%
%%%%%%%%%%%%%%%%%%%%%%%%%%%%%%%%%%%%%%%%%%%%%%%%%%%%%%%%%%%%%%%%%%%%%%%%%%%%%%%%

\begin{bibtex}[\jobname]

@article{Drummond:2009fd,
      author         = "Drummond, James M. and Henn, Johannes M. and Plefka, Jan",
      title          = "{Yangian symmetry of scattering amplitudes in N=4 super
                        Yang-Mills theory}",
      journal        = "JHEP",
      volume         = "0905",
      pages          = "046",
      doi            = "10.1088/1126-6708/2009/05/046",
      year           = "2009",
      eprint         = "0902.2987",
      archivePrefix  = "arXiv",
      primaryClass   = "hep-th",
      reportNumber   = "HU-EP-09-06, LAPTH-1308-09",
      SLACcitation   = "%%CITATION = ARXIV:0902.2987;%%",
}

@article{Drummond:2008vq,
      author         = "Drummond, J. M. and Henn, J. and Korchemsky, G. P. and
                        Sokatchev, E.",
      title          = "{Dual superconformal symmetry of scattering amplitudes in
                        N=4 super-Yang-Mills theory}",
      journal        = "Nucl.Phys.",
      volume         = "B828",
      pages          = "317-374",
      doi            = "10.1016/j.nuclphysb.2009.11.022",
      year           = "2010",
      eprint         = "0807.1095",
      archivePrefix  = "arXiv",
      primaryClass   = "hep-th",
      reportNumber   = "LAPTH-1257-08, LPT-ORSAY-08-60",
      SLACcitation   = "%%CITATION = ARXIV:0807.1095;%%",
}

@article{CaronHuot:2011kk,
      author         = "Caron-Huot, Simon and He, Song",
      title          = "{Jumpstarting the All-Loop S-Matrix of Planar N=4 Super
                        Yang-Mills}",
      journal        = "JHEP",
      volume         = "1207",
      pages          = "174",
      doi            = "10.1007/JHEP07(2012)174",
      year           = "2012",
      eprint         = "1112.1060",
      archivePrefix  = "arXiv",
      primaryClass   = "hep-th",
      SLACcitation   = "%%CITATION = ARXIV:1112.1060;%%",
}

@article{Bargheer:2011mm,
      author         = "Bargheer, Till and Beisert, Niklas and Loebbert, Florian",
      title          = "{Exact Superconformal and Yangian Symmetry of Scattering
                        Amplitudes}",
      journal        = "J.Phys.",
      volume         = "A44",
      pages          = "454012",
      doi            = "10.1088/1751-8113/44/45/454012",
      year           = "2011",
      eprint         = "1104.0700",
      archivePrefix  = "arXiv",
      primaryClass   = "hep-th",
      reportNumber   = "AEI-2011-016, LPT-ENS-11-12, UUITP-11-11",
      SLACcitation   = "%%CITATION = ARXIV:1104.0700;%%",
}

@article{Beisert:2010jq,
      author         = "Beisert, Niklas",
      title          = "{On Yangian Symmetry in Planar N=4 SYM}",
      year           = "2010",
      eprint         = "1004.5423",
      archivePrefix  = "arXiv",
      primaryClass   = "hep-th",
      reportNumber   = "AEI-2010-029",
      SLACcitation   = "%%CITATION = ARXIV:1004.5423;%%",
}

@article{Beisert:2010jr,
      author         = "Beisert, Niklas and others",
      title          = "{Review of AdS/CFT Integrability: An Overview}",
      journal        = "Lett.Math.Phys.",
      volume         = "99",
      pages          = "3-32",
      doi            = "10.1007/s11005-011-0529-2",
      year           = "2012",
      eprint         = "1012.3982",
      archivePrefix  = "arXiv",
      primaryClass   = "hep-th",
      reportNumber   = "AEI-2010-175, CERN-PH-TH-2010-306, HU-EP-10-87,
                        HU-MATH-2010-22, KCL-MTH-10-10, UMTG-270, UUITP-41-10",
      SLACcitation   = "%%CITATION = ARXIV:1012.3982;%%",
}

@article{Bargheer:2009qu,
      author         = "Bargheer, Till and Beisert, Niklas and Galleas,
                        Wellington and Loebbert, Florian and McLoughlin, Tristan",
      title          = "{Exacting N=4 Superconformal Symmetry}",
      journal        = "JHEP",
      volume         = "0911",
      pages          = "056",
      doi            = "10.1088/1126-6708/2009/11/056",
      year           = "2009",
      eprint         = "0905.3738",
      archivePrefix  = "arXiv",
      primaryClass   = "hep-th",
      reportNumber   = "AEI-2009-048",
      SLACcitation   = "%%CITATION = ARXIV:0905.3738;%%",
}

@article{Beisert:2010gn,
      author         = "Beisert, Niklas and Henn, Johannes and McLoughlin,
                        Tristan and Plefka, Jan",
      title          = "{One-Loop Superconformal and Yangian Symmetries of
                        Scattering Amplitudes in N=4 Super Yang-Mills}",
      journal        = "JHEP",
      volume         = "1004",
      pages          = "085",
      doi            = "10.1007/JHEP04(2010)085",
      year           = "2010",
      eprint         = "1002.1733",
      archivePrefix  = "arXiv",
      primaryClass   = "hep-th",
      reportNumber   = "AEI-2010-019, HU-EP-10-06",
      SLACcitation   = "%%CITATION = ARXIV:1002.1733;%%",
}

@article{Ferro:2012xw,
      author         = "Ferro, Livia and \L{}ukowski, Tomasz and Meneghelli, Carlo
                        and Plefka, Jan and Staudacher, Matthias",
      title          = "{Harmonic R-matrices for Scattering Amplitudes and
                        Spectral Regularization}",
      journal        = "Phys.Rev.Lett.",
      volume         = "110",
      pages          = "121602",
      doi            = "10.1103/PhysRevLett.110.121602",
      year           = "2013",
      eprint         = "1212.0850",
      archivePrefix  = "arXiv",
      primaryClass   = "hep-th",
      reportNumber   = "HU-EP-12-50, HU-MATHEMATIK:14-2012, DESY-12-228,
                        ZMP-HH-12-26, AEI-2012-198, --AEI-2012-198",
      SLACcitation   = "%%CITATION = ARXIV:1212.0850;%%",
}

@article{Ferro:2013dga,
      author         = "Ferro, Livia and Łukowski, Tomasz and Meneghelli, Carlo
                        and Plefka, Jan and Staudacher, Matthias",
      title          = "{Spectral Parameters for Scattering Amplitudes in N=4
                        Super Yang-Mills Theory}",
      journal        = "JHEP",
      volume         = "1401",
      pages          = "094",
      doi            = "10.1007/JHEP01(2014)094",
      year           = "2014",
      eprint         = "1308.3494",
      archivePrefix  = "arXiv",
      primaryClass   = "hep-th",
      reportNumber   = "HU-MATHEMATIK-2013-12, HU-EP-13-33, AEI-2013-235,
                        DESY-13-488, --ZMP-HH-13-15",
      SLACcitation   = "%%CITATION = ARXIV:1308.3494;%%",
}

@article{ArkaniHamed:2012nw,
      author         = "Arkani-Hamed, Nima and Bourjaily, Jacob L. and Cachazo,
                        Freddy and Goncharov, Alexander B. and Postnikov,
                        Alexander and Trnka, Jaroslav",
      title          = "{Scattering Amplitudes and the Positive Grassmannian}",
      year           = "2012",
      eprint         = "1212.5605",
      archivePrefix  = "arXiv",
      primaryClass   = "hep-th",
      SLACcitation   = "%%CITATION = ARXIV:1212.5605;%%",
}

@article{ArkaniHamed:2010kv,
      author         = "Arkani-Hamed, Nima and Bourjaily, Jacob L. and Cachazo,
                        Freddy and Caron-Huot, Simon and Trnka, Jaroslav",
      title          = "{The All-Loop Integrand For Scattering Amplitudes in
                        Planar N=4 SYM}",
      journal        = "JHEP",
      volume         = "1101",
      pages          = "041",
      doi            = "10.1007/JHEP01(2011)041",
      year           = "2011",
      eprint         = "1008.2958",
      archivePrefix  = "arXiv",
      primaryClass   = "hep-th",
      SLACcitation   = "%%CITATION = ARXIV:1008.2958;%%",
}

@article{ArkaniHamed:2010gh,
      author         = "Arkani-Hamed, Nima and Bourjaily, Jacob L. and Cachazo,
                        Freddy and Trnka, Jaroslav",
      title          = "{Local Integrals for Planar Scattering Amplitudes}",
      journal        = "JHEP",
      volume         = "1206",
      pages          = "125",
      doi            = "10.1007/JHEP06(2012)125",
      year           = "2012",
      eprint         = "1012.6032",
      archivePrefix  = "arXiv",
      primaryClass   = "hep-th",
      SLACcitation   = "%%CITATION = ARXIV:1012.6032;%%",
}

@article{ArkaniHamed:2009dn,
      author         = "Arkani-Hamed, Nima and Cachazo, Freddy and Cheung,
                        Clifford and Kaplan, Jared",
      title          = "{A Duality For The S Matrix}",
      journal        = "JHEP",
      volume         = "1003",
      pages          = "020",
      doi            = "10.1007/JHEP03(2010)020",
      year           = "2010",
      eprint         = "0907.5418",
      archivePrefix  = "arXiv",
      primaryClass   = "hep-th",
      SLACcitation   = "%%CITATION = ARXIV:0907.5418;%%",
}

@article{Drummond:2008cr,
      author         = "Drummond, J. M. and Henn, J. M.",
      title          = "{All tree-level amplitudes in N=4 SYM}",
      journal        = "JHEP",
      volume         = "0904",
      pages          = "018",
      doi            = "10.1088/1126-6708/2009/04/018",
      year           = "2009",
      eprint         = "0808.2475",
      archivePrefix  = "arXiv",
      primaryClass   = "hep-th",
      reportNumber   = "LAPTH-1267-08",
      SLACcitation   = "%%CITATION = ARXIV:0808.2475;%%",
}

@article{Britto:2004ap,
      author         = "Britto, Ruth and Cachazo, Freddy and Feng, Bo",
      title          = "{New recursion relations for tree amplitudes of gluons}",
      journal        = "Nucl.Phys.",
      volume         = "B715",
      pages          = "499-522",
      doi            = "10.1016/j.nuclphysb.2005.02.030",
      year           = "2005",
      eprint         = "hep-th/0412308",
      archivePrefix  = "arXiv",
      primaryClass   = "hep-th",
      SLACcitation   = "%%CITATION = HEP-TH/0412308;%%",
}

@article{Britto:2005fq,
      author         = "Britto, Ruth and Cachazo, Freddy and Feng, Bo and Witten,
                        Edward",
      title          = "{Direct proof of tree-level recursion relation in
                        Yang-Mills theory}",
      journal        = "Phys.Rev.Lett.",
      volume         = "94",
      pages          = "181602",
      doi            = "10.1103/PhysRevLett.94.181602",
      year           = "2005",
      eprint         = "hep-th/0501052",
      archivePrefix  = "arXiv",
      primaryClass   = "hep-th",
      SLACcitation   = "%%CITATION = HEP-TH/0501052;%%",
}

@article{Brandhuber:2008pf,
      author         = "Brandhuber, Andreas and Heslop, Paul and Travaglini,
                        Gabriele",
      title          = "{A Note on dual superconformal symmetry of the N=4 super
                        Yang-Mills S-matrix}",
      journal        = "Phys.Rev.",
      volume         = "D78",
      pages          = "125005",
      doi            = "10.1103/PhysRevD.78.125005",
      year           = "2008",
      eprint         = "0807.4097",
      archivePrefix  = "arXiv",
      primaryClass   = "hep-th",
      reportNumber   = "QMUL-PH-08-15",
      SLACcitation   = "%%CITATION = ARXIV:0807.4097;%%",
}

@article{ArkaniHamed:2008gz,
      author         = "Arkani-Hamed, Nima and Cachazo, Freddy and Kaplan, Jared",
      title          = "{What is the Simplest Quantum Field Theory?}",
      journal        = "JHEP",
      volume         = "1009",
      pages          = "016",
      doi            = "10.1007/JHEP09(2010)016",
      year           = "2010",
      eprint         = "0808.1446",
      archivePrefix  = "arXiv",
      primaryClass   = "hep-th",
      SLACcitation   = "%%CITATION = ARXIV:0808.1446;%%",
}

@article{Witten:2003nn,
      author         = "Witten, Edward",
      title          = "{Perturbative gauge theory as a string theory in twistor
                        space}",
      journal        = "Commun.Math.Phys.",
      volume         = "252",
      pages          = "189-258",
      doi            = "10.1007/s00220-004-1187-3",
      year           = "2004",
      eprint         = "hep-th/0312171",
      archivePrefix  = "arXiv",
      primaryClass   = "hep-th",
      SLACcitation   = "%%CITATION = HEP-TH/0312171;%%",
}

@article {Postnikov:math0609764,
      author         = "Alexander Postnikov",
      title          = "Total positivity, Grassmannians, and networks",
      year           = "2006",
      eprint         = "math/0609764",
      doi            = ""
}

@book{Mason:1991rf,
      author         = "Mason, L. J. and Woodhouse, N. M. J.",
      title      = "{Integrability, selfduality, and twistor theory}",
      publisher      = "Oxford University Press, Oxford, UK",
      editor         = "London Mathematical Society Monographs",
      year           = "1991",
      SLACcitation   = "%%CITATION = INSPIRE-328552;%%",
}

@article{CaronHuot:2010zt,
      author         = "Caron-Huot, Simon",
      title          = "{Loops and trees}",
      journal        = "JHEP",
      volume         = "1105",
      pages          = "080",
      doi            = "10.1007/JHEP05(2011)080",
      year           = "2011",
      eprint         = "1007.3224",
      archivePrefix  = "arXiv",
      primaryClass   = "hep-ph",
      SLACcitation   = "%%CITATION = ARXIV:1007.3224;%%",
}

@article{Drummond:2010qh,
      author         = "Drummond, J. M. and Ferro, L.",
      title          = "{Yangians, Grassmannians and T-duality}",
      journal        = "JHEP",
      volume         = "1007",
      pages          = "027",
      doi            = "10.1007/JHEP07(2010)027",
      year           = "2010",
      eprint         = "1001.3348",
      archivePrefix  = "arXiv",
      primaryClass   = "hep-th",
      reportNumber   = "LAPTH-001-10",
      SLACcitation   = "%%CITATION = ARXIV:1001.3348;%%",
}

@article{Drummond:2010uq,
      author         = "Drummond, J. M. and Ferro, L.",
      title          = "{The Yangian origin of the Grassmannian integral}",
      journal        = "JHEP",
      volume         = "1012",
      pages          = "010",
      doi            = "10.1007/JHEP12(2010)010",
      year           = "2010",
      eprint         = "1002.4622",
      archivePrefix  = "arXiv",
      primaryClass   = "hep-th",
      SLACcitation   = "%%CITATION = ARXIV:1002.4622;%%",
}

@article{Bargheer:2012cp,
      author         = "Bargheer, Till and Beisert, Niklas and Loebbert, Florian
                        and McLoughlin, Tristan",
      title          = "{Conformal Anomaly for Amplitudes in N=6
                        Superconformal Chern-Simons Theory}",
      journal        = "J.Phys.",
      volume         = "A45",
      pages          = "475402",
      doi            = "10.1088/1751-8113/45/47/475402",
      year           = "2012",
      eprint         = "1204.4406",
      archivePrefix  = "arXiv",
      primaryClass   = "hep-th",
      reportNumber   = "NSF-KITP-12-012, LPT-ENS-12-16, UUITP-10-12,
                        AEI-2012-037",
      SLACcitation   = "%%CITATION = ARXIV:1204.4406;%%",
}

@article{Alday:2008yw,
      author         = "Alday, Luis F. and Roiban, Radu",
      title          = "{Scattering Amplitudes, Wilson Loops and the String/Gauge
                        Theory Correspondence}",
      journal        = "Phys.Rept.",
      volume         = "468",
      pages          = "153-211",
      doi            = "10.1016/j.physrep.2008.08.002",
      year           = "2008",
      eprint         = "0807.1889",
      archivePrefix  = "arXiv",
      primaryClass   = "hep-th",
      SLACcitation   = "%%CITATION = ARXIV:0807.1889;%%",
}

@article{Roiban:2011zz,
  author = {Roiban, R. and Spradlin, M. and Volovich, (eds.), A.},
  title = {Scattering Amplitudes in Gauge Theories: Progress and Outlook},
  journal = {J. Phys. A},
  volume = {44},
  pages = {450301},
  year = {2011},
  doi = {10.1088/1751-8113/44/45/450301},
}

@article{Bern:2008ap,
      author         = "Bern, Z. and Dixon, L.J. and Kosower, D.A. and Roiban, R.
                        and Spradlin, M. and others",
      title          = "{The Two-Loop Six-Gluon MHV Amplitude in Maximally
                        Supersymmetric Yang-Mills Theory}",
      journal        = "Phys.Rev.",
      volume         = "D78",
      pages          = "045007",
      doi            = "10.1103/PhysRevD.78.045007",
      year           = "2008",
      eprint         = "0803.1465",
      archivePrefix  = "arXiv",
      primaryClass   = "hep-th",
      reportNumber   = "SLAC-PUB-13150, SACLAY-IPHT-T08-045, UCLA-08-TEP-5,
                        BROWN-HET-1495",
      SLACcitation   = "%%CITATION = ARXIV:0803.1465;%%",
}

\end{bibtex}

\bibliographystyle{nb}
\bibliography{\jobname}

%bibliography generated by nb.bst v1.06 (C) 2003-2011 Niklas Beisert
\begin{thebibliography}{10}
\providecommand{\href}[2]{#2}
\providecommand{\arxivref}[2]{\href{http://arxiv.org/abs/#1}{#2}}
\providecommand{\doiref}[2]{\href{http://dx.doi.org/#1}{#2}}
\providecommand{\nbbstauthor}[1]{#1}
\providecommand{\nbbstjournal}[1]{\textsf{#1}}
\providecommand{\nbbsttitle}[1]{\textit{#1}}
\providecommand{\nbbsturl}[1]{\texttt{#1}}
\providecommand{\nbbsteprint}[1]{\texttt{#1}}
\providecommand{\nbbststyle}{\raggedright\small\parskip0pt}
\nbbststyle

\bibitem{Drummond:2009fd}
\nbbstauthor{J.~M.~Drummond, J.~M.~Henn and J.~Plefka},
\nbbsttitle{``{Yangian symmetry of scattering amplitudes in N=4 super
  Yang-Mills theory}''},
\nbbstjournal{\doiref{10.1088/1126-6708/2009/05/046}{JHEP~0905,~046~(2009)}},
\nbbsteprint{\arxivref{0902.2987}{arxiv:0902.2987}}.
%%CITATION = ARXIV:0902.2987;%%

\bibitem{Roiban:2011zz}
\nbbstauthor{R.~Roiban, M.~Spradlin and A.~Volovich,~(eds.)},
\nbbsttitle{``Scattering Amplitudes in Gauge Theories: Progress and Outlook''},
\nbbstjournal{\doiref{10.1088/1751-8113/44/45/450301}{J.~Phys.~A~44,~450301~(2%
011)}}.

\bibitem{Beisert:2010jq}
\nbbstauthor{N.~Beisert},
\nbbsttitle{``{On Yangian Symmetry in Planar N=4 SYM}''},
\nbbsteprint{\arxivref{1004.5423}{arxiv:1004.5423}}.
%%CITATION = ARXIV:1004.5423;%%

\bibitem{Drummond:2008vq}
\nbbstauthor{J.~M.~Drummond, J.~Henn, G.~P.~Korchemsky and E.~Sokatchev},
\nbbsttitle{``{Dual superconformal symmetry of scattering amplitudes in N=4
  super-Yang-Mills theory}''},
\nbbstjournal{\doiref{10.1016/j.nuclphysb.2009.11.022}{Nucl.~Phys.~B828,~317~(%
2010)}},
\nbbsteprint{\arxivref{0807.1095}{arxiv:0807.1095}}.
%%CITATION = ARXIV:0807.1095;%%

\bibitem{Beisert:2010jr}
\nbbstauthor{N.~Beisert et~al.},
\nbbsttitle{``{Review of AdS/CFT Integrability: An Overview}''},
\nbbstjournal{\doiref{10.1007/s11005-011-0529-2}{Lett.~Math.~Phys.~99,~3~(2012%
)}},
\nbbsteprint{\arxivref{1012.3982}{arxiv:1012.3982}}.
%%CITATION = ARXIV:1012.3982;%%

\bibitem{ArkaniHamed:2012nw}
\nbbstauthor{N.~Arkani-Hamed, J.~L.~Bourjaily, F.~Cachazo, A.~B.~Goncharov,
  A.~Postnikov and J.~Trnka},
\nbbsttitle{``{Scattering Amplitudes and the Positive Grassmannian}''},
\nbbsteprint{\arxivref{1212.5605}{arxiv:1212.5605}}.
%%CITATION = ARXIV:1212.5605;%%

\bibitem{ArkaniHamed:2009dn}
\nbbstauthor{N.~Arkani-Hamed, F.~Cachazo, C.~Cheung and J.~Kaplan},
\nbbsttitle{``{A Duality For The S Matrix}''},
\nbbstjournal{\doiref{10.1007/JHEP03(2010)020}{JHEP~1003,~020~(2010)}},
\nbbsteprint{\arxivref{0907.5418}{arxiv:0907.5418}}.
%%CITATION = ARXIV:0907.5418;%%

\bibitem{Drummond:2010qh}
\nbbstauthor{J.~M.~Drummond and L.~Ferro},
\nbbsttitle{``{Yangians, Grassmannians and T-duality}''},
\nbbstjournal{\doiref{10.1007/JHEP07(2010)027}{JHEP~1007,~027~(2010)}},
\nbbsteprint{\arxivref{1001.3348}{arxiv:1001.3348}}.
%%CITATION = ARXIV:1001.3348;%%

\bibitem{Drummond:2010uq}
\nbbstauthor{J.~M.~Drummond and L.~Ferro},
\nbbsttitle{``{The Yangian origin of the Grassmannian integral}''},
\nbbstjournal{\doiref{10.1007/JHEP12(2010)010}{JHEP~1012,~010~(2010)}},
\nbbsteprint{\arxivref{1002.4622}{arxiv:1002.4622}}.
%%CITATION = ARXIV:1002.4622;%%

\bibitem{ArkaniHamed:2010kv}
\nbbstauthor{N.~Arkani-Hamed, J.~L.~Bourjaily, F.~Cachazo, S.~Caron-Huot and
  J.~Trnka},
\nbbsttitle{``{The All-Loop Integrand For Scattering Amplitudes in Planar N=4
  SYM}''},
\nbbstjournal{\doiref{10.1007/JHEP01(2011)041}{JHEP~1101,~041~(2011)}},
\nbbsteprint{\arxivref{1008.2958}{arxiv:1008.2958}}.
%%CITATION = ARXIV:1008.2958;%%

\bibitem{ArkaniHamed:2010gh}
\nbbstauthor{N.~Arkani-Hamed, J.~L.~Bourjaily, F.~Cachazo and J.~Trnka},
\nbbsttitle{``{Local Integrals for Planar Scattering Amplitudes}''},
\nbbstjournal{\doiref{10.1007/JHEP06(2012)125}{JHEP~1206,~125~(2012)}},
\nbbsteprint{\arxivref{1012.6032}{arxiv:1012.6032}}.
%%CITATION = ARXIV:1012.6032;%%

\bibitem{Ferro:2012xw}
\nbbstauthor{L.~Ferro, T.~\L{}ukowski, C.~Meneghelli, J.~Plefka and
  M.~Staudacher},
\nbbsttitle{``{Harmonic R-matrices for Scattering Amplitudes and Spectral
  Regularization}''},
\nbbstjournal{\doiref{10.1103/PhysRevLett.110.121602}{Phys.~Rev.~Lett.~110,~12%
1602~(2013)}},
\nbbsteprint{\arxivref{1212.0850}{arxiv:1212.0850}}.
%%CITATION = ARXIV:1212.0850;%%

\bibitem{Ferro:2013dga}
\nbbstauthor{L.~Ferro, T.~Łukowski, C.~Meneghelli, J.~Plefka and
  M.~Staudacher},
\nbbsttitle{``{Spectral Parameters for Scattering Amplitudes in N=4 Super
  Yang-Mills Theory}''},
\nbbstjournal{\doiref{10.1007/JHEP01(2014)094}{JHEP~1401,~094~(2014)}},
\nbbsteprint{\arxivref{1308.3494}{arxiv:1308.3494}}.
%%CITATION = ARXIV:1308.3494;%%

\bibitem{Britto:2004ap}
\nbbstauthor{R.~Britto, F.~Cachazo and B.~Feng},
\nbbsttitle{``{New recursion relations for tree amplitudes of gluons}''},
\nbbstjournal{\doiref{10.1016/j.nuclphysb.2005.02.030}{Nucl.~Phys.~B715,~499~(%
2005)}},
\nbbsteprint{\arxivref{hep-th/0412308}{hep-th/0412308}}.
%%CITATION = HEP-TH/0412308;%%

\bibitem{Britto:2005fq}
\nbbstauthor{R.~Britto, F.~Cachazo, B.~Feng and E.~Witten},
\nbbsttitle{``{Direct proof of tree-level recursion relation in Yang-Mills
  theory}''},
\nbbstjournal{\doiref{10.1103/PhysRevLett.94.181602}{Phys.~Rev.~Lett.~94,~1816%
02~(2005)}},
\nbbsteprint{\arxivref{hep-th/0501052}{hep-th/0501052}}.
%%CITATION = HEP-TH/0501052;%%

\bibitem{ArkaniHamed:2008gz}
\nbbstauthor{N.~Arkani-Hamed, F.~Cachazo and J.~Kaplan},
\nbbsttitle{``{What is the Simplest Quantum Field Theory?}''},
\nbbstjournal{\doiref{10.1007/JHEP09(2010)016}{JHEP~1009,~016~(2010)}},
\nbbsteprint{\arxivref{0808.1446}{arxiv:0808.1446}}.
%%CITATION = ARXIV:0808.1446;%%

\bibitem{Brandhuber:2008pf}
\nbbstauthor{A.~Brandhuber, P.~Heslop and G.~Travaglini},
\nbbsttitle{``{A Note on dual superconformal symmetry of the N=4 super
  Yang-Mills S-matrix}''},
\nbbstjournal{\doiref{10.1103/PhysRevD.78.125005}{Phys.~Rev.~D78,~125005~(2008%
)}},
\nbbsteprint{\arxivref{0807.4097}{arxiv:0807.4097}}.
%%CITATION = ARXIV:0807.4097;%%

\bibitem{Mason:1991rf}
\nbbstauthor{L.~J.~Mason and N.~M.~J.~Woodhouse},
\nbbsttitle{``{Integrability, selfduality, and twistor theory}''},
Oxford University Press, Oxford, UK (1991).

\bibitem{Witten:2003nn}
\nbbstauthor{E.~Witten},
\nbbsttitle{``{Perturbative gauge theory as a string theory in twistor
  space}''},
\nbbstjournal{\doiref{10.1007/s00220-004-1187-3}{Commun.~Math.~Phys.~252,~189~%
(2004)}},
\nbbsteprint{\arxivref{hep-th/0312171}{hep-th/0312171}}.
%%CITATION = HEP-TH/0312171;%%

\bibitem{Bargheer:2009qu}
\nbbstauthor{T.~Bargheer, N.~Beisert, W.~Galleas, F.~Loebbert and
  T.~McLoughlin},
\nbbsttitle{``{Exacting N=4 Superconformal Symmetry}''},
\nbbstjournal{\doiref{10.1088/1126-6708/2009/11/056}{JHEP~0911,~056~(2009)}},
\nbbsteprint{\arxivref{0905.3738}{arxiv:0905.3738}}.
%%CITATION = ARXIV:0905.3738;%%

\bibitem{Bargheer:2011mm}
\nbbstauthor{T.~Bargheer, N.~Beisert and F.~Loebbert},
\nbbsttitle{``{Exact Superconformal and Yangian Symmetry of Scattering
  Amplitudes}''},
\nbbstjournal{\doiref{10.1088/1751-8113/44/45/454012}{J.~Phys.~A44,~454012~(20%
11)}},
\nbbsteprint{\arxivref{1104.0700}{arxiv:1104.0700}}.
%%CITATION = ARXIV:1104.0700;%%

\bibitem{Bargheer:2012cp}
\nbbstauthor{T.~Bargheer, N.~Beisert, F.~Loebbert and T.~McLoughlin},
\nbbsttitle{``{Conformal Anomaly for Amplitudes in N=6 Superconformal
  Chern-Simons Theory}''},
\nbbstjournal{\doiref{10.1088/1751-8113/45/47/475402}{J.~Phys.~A45,~475402~(20%
12)}},
\nbbsteprint{\arxivref{1204.4406}{arxiv:1204.4406}}.
%%CITATION = ARXIV:1204.4406;%%

\bibitem{CaronHuot:2010zt}
\nbbstauthor{S.~Caron-Huot},
\nbbsttitle{``{Loops and trees}''},
\nbbstjournal{\doiref{10.1007/JHEP05(2011)080}{JHEP~1105,~080~(2011)}},
\nbbsteprint{\arxivref{1007.3224}{arxiv:1007.3224}}.
%%CITATION = ARXIV:1007.3224;%%

\bibitem{Bern:2008ap}
\nbbstauthor{Z.~Bern, L.~Dixon, D.~Kosower, R.~Roiban, M.~Spradlin et~al.},
\nbbsttitle{``{The Two-Loop Six-Gluon MHV Amplitude in Maximally Supersymmetric
  Yang-Mills Theory}''},
\nbbstjournal{\doiref{10.1103/PhysRevD.78.045007}{Phys.~Rev.~D78,~045007~(2008%
)}},
\nbbsteprint{\arxivref{0803.1465}{arxiv:0803.1465}}.
%%CITATION = ARXIV:0803.1465;%%

\bibitem{Postnikov:math0609764}
\nbbstauthor{A.~Postnikov},
\nbbsttitle{``Total positivity, Grassmannians, and networks''},
\nbbsteprint{\arxivref{math/0609764}{math/0609764}}.

\end{thebibliography}

\end{document}